\documentclass[amsmath,amssymb,showkeys,superscriptaddress,floatfix,aps,prd,10pt]{revtex4}
\usepackage{graphicx,epstopdf}
\usepackage[caption=false]{subfig}
\usepackage{multirow}
\usepackage{appendix}
\usepackage{xcolor}
\usepackage[colorlinks=true, urlcolor=blue, linkcolor=blue, citecolor=green]{hyperref}
\def\pslash{p\!\!\!\slash }
\def\qslash{q\!\!\!\slash }
\def\xslash{x\!\!\!\slash }

\def\pslash{p\!\!\!\slash }
\def\qslash{q\!\!\!\slash }
\def\xslash{x\!\!\!\slash }

\def\beq{\begin{equation}}
\def\eeq{\end{equation}}
\def\bea{\begin{eqnarray}}
\def\eea{\end{eqnarray}}
\def\beeq{\begin{eqnarray}}
\def\eeeq{\end{eqnarray}}

\def\ba{\begin{array}}
\def\ea{\end{array}}

\def\xis0{{\Xi^{*0}}}

\def\g5{\gamma_5}

\begin{document}

\title{Gravitational transition form factors of $N \rightarrow \Delta$ via QCD light-cone sum rules}

\author{U. \"{O}zdem}
\affiliation{Health Services Vocational School of Higher Education, Istanbul Aydin University, Sefakoy-Kucukcekmece, 34295 Istanbul, T\"{u}rkiye}
\author{K. Azizi}
\affiliation{Department of Physics, University of Tehran, North Karegar Avenue, Tehran
14395-547, Iran}
\affiliation{Department of Physics, Do\v{g}u\c{s} University, Dudullu-\"{U}mraniye, 34775
Istanbul,  T\"{u}rkiye}

\begin{abstract}
We present the first direct calculation on the gravitational  form factors (GFFs)  of  the $N \rightarrow \Delta$  transition using an analytic method,  the QCD light-cone sum rules. The matrix element of the quark part of the energy momentum tensor current sandwiched between the nucleon and $\Delta$ states are parameterized in terms of five independent conserved and four independent non-conserved  GFFs, for calculation of which we use the distribution amplitudes (DAs) of the on-shell nucleon expanded in terms of  functions with different twists. We present the results for two sets of light-cone input parameters. The results indicate that the behavior of the form factors with respect to  $Q^2$ are described by multipole fit functions. Our results may be checked by other phenomenological  models including the Lattice QCD as well as future related experiments.
\end{abstract}
 \date{\today}
\maketitle

\section{Introduction} 
The interaction of composite particles with electromagnetic, QCD energy-momentum and weak currents are parameterized in terms of the corresponding form factors (FFs) as the building blocks. Having determined these FFs, one can construct different observables in terms of these non-perturbative objects to determine the nature and internal quark-gluon organizations of hadrons. In the case of  the energy-momentum tensor (EMT) or  gravitational form factors (GFFs), which represent the gravitational interaction between a graviton and  the hadron,    these FFs provide information on the  mechanical properties of the hadron  such as the mass, spin, pressure,  shear force, radius, etc. Understanding  the mechanical structure of hadrons   is of  great importance as it  provides us with the  fundamental information on the internal structure and geometric shapes  of the hadrons.

The GFFs for the spin-1/2 particles entering the matrix elements of the
EMT current were first parameterized in Refs. \cite{Kobzarev:1962wt,Pagels:1966zza,Kobsarev:1970qm,Ng:1993vh}. Based on these  parametrizations, the GFFs of the nucleon as a basic  baryonic  structure have been intensively
investigated in various approaches~\cite{Polyakov:2002wz,Ji:1997gm,Schweitzer:2002nm,Jung:2013bya,Hagler:2003jd,Gockeler:2003jfa,Pasquini:2007xz,Hwang:2007tb,Abidin:2008hn,Brodsky:2008pf,Pasquini:2014vua,Chakrabarti:2015lba,Lorce:2018egm,Teryaev:2016edw,Shanahan:2018nnv, Shanahan:2018pib,Neubelt:2019sou,Anikin:2019kwi,Alharazin:2020yjv,Gegelia:2021wnj,Varma:2020crx,Fujita:2022jus,Mamo:2022eui,Pefkou:2021fni,Azizi:2019ytx,Polyakov:2018exb,Freese:2021mzg,Freese:2021qtb,Freese:2021czn, Burkert:2021ith,Burkert:2018bqq}. Compared to the nucleon, the GFFs of other members of the  baryon octet were much less studied~\cite{Ozdem:2020ieh,Won:2022cyy}.
The calculations of EMT FFs  have recently  been extended to the parity flip   transitions, $ N^*\rightarrow N $~\cite{Polyakov:2020rzq, Azizi:2020jog, Ozdem:2019pkg}.    They were also generalized to the higher-spin particles~\cite{Cotogno:2019vjb} in a systematic way.  For a spin-1 particle, the corresponding GFFs  were studied in Refs.~\cite{Polyakov:2019lbq,Cosyn:2019aio,Kim:2022wkc,Freese:2022yur,Freese:2022ibw,Freese:2019bhb,Sun:2020wfo,Epelbaum:2021ahi,Pefkou:2021fni} using different models and approaches. 
Very recently, the $N \to \Delta$ transition matrix elements of the EMT current were also parametrized in Ref. \cite{Kim:2022bwn}.
The GFFs for the spin-3/2 particle were also examined in Refs. \cite{Pefkou:2021fni,Fu:2022rkn,Alharazin:2022wjj,Panteleeva:2020ejw}, recently. 

In our study on the GFFs of the nucleon \cite{Azizi:2019ytx}, which  were calculated in a large range of transferred momentum square $ Q^2 $, we compared our predictions on  the GFFs  $M_2^q(Q^2)$, $J^q(Q^2)$ and $d_1^q(Q^2)$  (D-term) at small values of $ Q^2 $ with the existing Lattice QCD  predictions \cite{Hagler:2007xi}  as well as the JLab data for the D-term \cite{Burkert:2018bqq}.  For the form factor $M_2^q(Q^2)$, our predictions were consistent with most of the Lattice QCD data points considering the errorbars. In the case of $J^q(Q^2)$ and $d_1^q(Q^2)$ form factors,  the Lattice results suffer from large uncertainties at small values of $ Q^2 $ that should be computed more accurately. It was also obtained that our predictions reproduce most of the JLab data at small  values of $ Q^2 $. 
In general, due to polynomiality, all the GFFs of the nucleon can also be obtained from the second  Mellin moments of the generalized parton distributions (GPDs) integrated over  Bjorken-$ x $, for any value of the skewness variable $\xi$. Only, the $M_2(Q^2)$ and $J(Q^2)$ GFFs of the  nucleon can  be  calculated  using  its GPDs at zero skewness \cite {Polyakov:2002yz}. Very recently, the $M_2(Q^2)$ and $J(Q^2)$ GFFs were obtained  at this limit  by integration of  the GPDs $H^q(x,Q^2)  $ and  $E^q(x,Q^2)  $ over  Bjorken-$ x $ in Refs. \cite{ Hashamipour:2022noy,Hashamipour:2021kes} and compared with our previous light cone QCD sum rules predictions  presented in Ref. \cite{Azizi:2019ytx}.  Very surprisingly, for special sets,  the obtained results via the GPDs that themselves are extracted from the pure experimental data of different collaborations are in a nice consistency with the predictions of our previous study,   Ref. \cite{Azizi:2019ytx},  for  the GFFs $M_2(Q^2)$ and $J(Q^2)$ at a vide range of $ Q^2 $.  Our predictions on the GFFs of the negative parity excited nucleon $N(1535) $  or  $N^* $ and the transition GFFs of  the  $ N^*\rightarrow N $ \cite{ Azizi:2020jog, Ozdem:2019pkg} can be used to be compared with the future Lattice QCD results as well as experimental data.

In the present article, we study the transition GFFs of the $N \to \Delta$ for the first time. We use the light cone QCD sum rule formalism to find  the five independent conserved and four independent non-conserved  GFFs defining this transition by considering the nucleon as the on-shell state, which allows us to use the DAs of the nucleon in terms of wavefunctions of different twists. We use two sets of parameters entering the wavefunctions to numerically analyze the obtained sum rules for the FFs. We also give the fit functions of the transition form factors  describing the behavior of the FFs in terms of $ Q^2 $.

The article is structured as follows. In next section we derive the desired sum rules for the transition GFFs of  the $N \to \Delta$. In section III, we numerically analyze the sum rules for the  form factors to find their  $ Q^2 $ 		behavior and their values at static limit . The last section is reserved for our concluding remarks.

\section{Formalism} \label{secII}
In light-cone  sum rule (LCSR) approach, the following two-point correlation function in the presence of  the on-shell nucleon state is responsible for the calculations of the GFFs of the $N \to \Delta$ transition:
 \begin{equation}\label{corf}
	\Pi_{\alpha\mu\nu}(p,q)=i\int d^4 x e^{iqx} \langle 0 |\mathcal{T}[J_{\alpha}^{\Delta}(0)T_{\mu \nu}^q(x)]|N(p)\rangle,
\end{equation}
where $ \mathcal{T} $ is the time ordering operator,  $p$ is the nucleon's four-momentum, $q$ is the transferred momentum, 
 $J_{\alpha}^{\Delta}(0)$ is  the $\Delta$'s interpolating  current and    $T_{\mu\nu}^q(x)$ is the quark part of the EMT current at point $ x $.  We insert a complete set of the intermediate $\Delta(p',s')  $ with momentum $ p' $ and $ ,s' $ to the correlation function and perform the four-integral over $ x $. This ends up in
 \begin{align}\label{phys}
 \Pi_{\alpha\mu\nu}^{Had}(p,q) &=\sum_{s{'}} \frac{\langle0|J_{\alpha}^{\Delta}|{\Delta(p',s')}\rangle\langle {\Delta(p',s')}
 |T_{\mu \nu}^q|N(p,s)\rangle}{m^2_{\Delta}-p'^2} 
 +...,
\end{align}
where dots stand for the contributions of the higher states and continuum. We choose the continuum threshold $ s_0 $ (coming from the continuum subtraction procedure)  such that the correlation function includes only the ground state $\Delta$ and  the first and higher excited states are included in dots in the above equation (for the procedure of the continuum subtraction and determination of the working window for continuum threshold $ s_0 $ see the sections Appendix and   III, respectively). To go further, we need to define the following matrix element in terms of  the residue of the $\Delta$ baryon ($ \lambda_\Delta $):
 \begin{align}\label{resi1}
 \langle0|J_{\alpha}^{\Delta}|{\Delta(p',s')}\rangle &= \lambda_{\Delta} u_{\alpha}(p',s'),
  \end{align}
   where,  $ u_{\alpha}(p',s') $   is the Rarita-Schwinger spinor. To proceed, we define the  matrix element of the quark part of the EMT current sandwiched between the nucleon and $\Delta$ state. As the quark part of the EMT current  solely is not a conserved quantity, the transition matrix element is decomposed in terms of nine form factors (five independent conserved and four independent non-conserved) by demanding the criteria of the  Lorentz invariance, discrete space-time symmetries and the related equations of motions  \cite{Kim:2022bwn}: 
\begin{widetext}
 \begin{align}\label{GFFs}
 \langle \Delta(p^\prime,s')|T_{\mu\nu}^q|N(p,s)\rangle &=
 \bar{u}_{\beta}(p^\prime,s')
 \Bigg[F_1^{N\Delta}(Q^2)\bigg\{g_{\beta\{\mu}P_{\nu\}}+\frac{(m^2_{\Delta}-m^2_N)}{\Delta^2}g_{\mu\nu} \Delta_{\beta}- \frac{(m^2_{\Delta}-m^2_N)}{2 \Delta^2}g_{\beta\{ \mu}\Delta_{\nu\}}-\frac{\Delta_\beta P_{\{\mu} \Delta_{\nu\}}}{\Delta^2}
     \bigg\}\nonumber\\
 &+\frac{F_2^{N\Delta}(Q^2)}{{\bar m}^2}\bigg\{ P_\mu P_\nu \Delta_\beta + \frac{(m_\Delta^2-m_N^2)^2}{4 \Delta^2}g_{\mu\nu} \Delta_\beta- \frac{(m^2_{\Delta}-m^2_N)}{2 \Delta^2}P_{\{\mu}\Delta_{\nu\}}\Delta_\beta
 \bigg\}\nonumber\\
 &+\frac{F_3^{N\Delta}(Q^2)}{{\bar m}^2}\bigg\{ (\Delta_{\mu}\Delta_{\nu}- \Delta^2  g_{\mu\nu})\Delta_\beta \bigg\}
 \nonumber\\
 &+F_4^{N\Delta}(Q^2) \bar m \bigg\{ \gamma_{\{\mu} g_{ \nu\}\beta} + \frac{2(m_\Delta+m_N)}{\Delta^2} g_{\mu\nu} \Delta_\beta  -    \frac{(m_\Delta+m_N)}{\Delta^2} g_{\beta\{\mu}\Delta_{\nu\}}-\frac{1}{\Delta^2}  \gamma_{\{\mu}\Delta_{\nu\}}\Delta_\beta                  
 \bigg\}\nonumber\\
 &+\frac{ F_5^{N\Delta}(Q^2)}{\bar m}\bigg\{   \gamma_{\{\mu}P_{\nu\}}\Delta_\beta   +\frac{(m^2_{\Delta}-m^2_N)(m_\Delta+m_N)}{\Delta^2}  g_{\mu\nu}\Delta_\beta   -     \frac{(m_\Delta+m_N)}{\Delta^2}\Delta_{\{\mu}P_{\nu\}}\Delta_\beta  \nonumber\\
 &-  \frac{(m^2_{\Delta}-m^2_N)}{2\Delta^2}\gamma_{\{\mu}\Delta_{\nu\}}\Delta_\beta
 \bigg\}
 + \bar C_1^{N\Delta}  (Q^2)  g_{\mu\nu}\Delta_\beta 
 +\frac{\bar C_2^{N\Delta}  (Q^2)}{{\bar m}^2} \Delta_{\{\mu} P_{\nu\}}\Delta_\beta  
 \nonumber\\
  &
  +\frac{\bar C_3^{N\Delta} (Q^2)}{{\bar m}}  \gamma_{\{\mu}\Delta_{\nu\}} \Delta_\beta
  +\bar C_4^{N\Delta} (Q^2) g_{\beta\{ \mu}\Delta_{\nu\}}
  \Bigg] \gamma_5 u_{N}(p,s),
 \end{align}
 
\end{widetext}
where $ P= (p'+p)/2$, 
$\Delta = p'-p$,  $\bar m = (m_N + m_\Delta)/2$, 
$X_{\{\mu}Y_{\nu\}} = (X_{\mu}Y_{\nu}+ X_{\nu}Y_{\mu})/2$ and  $Q^2=- \Delta^2$.
Here, $F_1^{N\Delta}(Q^2)  $, $F_2^{N\Delta}(Q^2) $, $F_3^{N\Delta}(Q^2) $, $F_4^{N\Delta}(Q^2) $, $F_5^{N\Delta}(Q^2) $,   $\bar C_1^{N\Delta} (Q^2) $, $\bar C_2^{N\Delta} (Q^2) $, $\bar C_3^{N\Delta} (Q^2) $, and $\bar C_4^{N\Delta} (Q^2) $ are the transition GFFs.  Note that by introducing $\bar m$ into the above definition at different places we tried to bring all the form factors to the  same dimensions.
 To further simplify,  the  summation over the spin of the  the Rarita-Schwinger spinor for the $\Delta$ baryon is introduced:
\begin{equation}\label{sspin}
\sum_{s'}{u_\alpha}(p',s'){\bar u_{\beta}}(p',s')=-(\not\!p'+m_{\Delta})\{g_{\alpha\beta}-\frac{1}{3}\gamma_{\alpha}\gamma_{\beta}-\frac{2p'_{\alpha}p'_{\beta}}
{3m_{\Delta}^{2}}+\frac{p'_{\alpha}\gamma_{\beta}-p'_{\beta}\gamma_{\alpha}}{3m_{\Delta}}\}.
\end{equation}
Using  Eqs. (\ref{resi1}), \eqref{GFFs} and \eqref{sspin} in Eq. \eqref{phys},  we recast the phenomenological or physical  form of the correlation function in terms of  the  GFFs and other corresponding  hadronic  parameters  as: 
\begin{widetext}
 \begin{align}\label{physson}
\Pi_{\alpha \mu\nu}^{Had} (p,q)&= \frac{\lambda_{\Delta}}{m^2_{\Delta}-p'^2}\Bigg[ -(\not\!p'+m_{\Delta})\{g_{\alpha\beta}-\frac{1}{3}\gamma_{\alpha}\gamma_{\beta}-\frac{2p'_{\alpha}p'_{\beta}}
{3m_{\Delta}^{2}}+\frac{p'_{\alpha}\gamma_{\beta}-p'_{\beta}\gamma_{\alpha}}{3m_{\Delta}}\}\Bigg]\nonumber\\
&\times \Bigg[F_1^{N\Delta}(Q^2)\bigg\{g_{\beta\{\mu}P_{\nu\}}+\frac{(m^2_{\Delta}-m^2_N)}{\Delta^2}g_{\mu\nu} \Delta_{\beta}- \frac{(m^2_{\Delta}-m^2_N)}{2 \Delta^2}g_{\beta\{ \mu}\Delta_{\nu\}}-\frac{\Delta_\beta P_{\{\mu} \Delta_{\nu\}}}{\Delta^2}
     \bigg\}\nonumber\\
 &+\frac{F_2^{N\Delta}(Q^2)}{{\bar m}^2}\bigg\{ P_\mu P_\nu \Delta_\beta + \frac{(m_\Delta^2-m_N^2)^2}{4 \Delta^2}g_{\mu\nu} \Delta_\beta- \frac{(m^2_{\Delta}-m^2_N)}{2 \Delta^2}P_{\{\mu}\Delta_{\nu\}}\Delta_\beta
 \bigg\}\nonumber\\
 &+\frac{F_3^{N\Delta}(Q^2)}{{\bar m}^2}\bigg\{ (\Delta_{\mu}\Delta_{\nu}- \Delta^2  g_{\mu\nu})\Delta_\beta \bigg\}
 \nonumber\\
 &+F_4^{N\Delta}(Q^2) \bar m \bigg\{ \gamma_{\{\mu} g_{ \nu\}\beta} + \frac{2(m_\Delta+m_N)}{\Delta^2} g_{\mu\nu} \Delta_\beta  -    \frac{(m_\Delta+m_N)}{\Delta^2} g_{\beta\{\mu}\Delta_{\nu\}}-\frac{1}{\Delta^2}  \gamma_{\{\mu}\Delta_{\nu\}}\Delta_\beta                  
 \bigg\}\nonumber\\
 &+\frac{ F_5^{N\Delta}(Q^2)}{\bar m}\bigg\{   \gamma_{\{\mu}P_{\nu\}}\Delta_\beta   +\frac{(m^2_{\Delta}-m^2_N)(m_\Delta+m_N)}{\Delta^2}  g_{\mu\nu}\Delta_\beta   -     \frac{(m_\Delta+m_N)}{\Delta^2}\Delta_{\{\mu}P_{\nu\}}\Delta_\beta  \nonumber\\
 &-  \frac{(m^2_{\Delta}-m^2_N)}{2\Delta^2}\gamma_{\{\mu}\Delta_{\nu\}}\Delta_\beta
 \bigg\}
 + \bar C_1^{N\Delta}  (Q^2)  g_{\mu\nu}\Delta_\beta 
 +\frac{\bar C_2^{N\Delta}  (Q^2)}{{\bar m}^2} \Delta_{\{\mu} P_{\nu\}}\Delta_\beta  
 \nonumber\\
  &
  +\frac{\bar C_3^{N\Delta} (Q^2)}{{\bar m}}  \gamma_{\{\mu}\Delta_{\nu\}} \Delta_\beta
  +\bar C_4^{N\Delta} (Q^2) g_{\beta\{ \mu}\Delta_{\nu\}}
  \Bigg] \gamma_5 u_{N}(p,s),
\end{align}
where we only kept  the contribution of the spin-$3/2$ $\Delta$ baryon  and omitted the contamination coming from the spin-$1/2$ particles. 
In principle, the correlation function can also include the contributions from spin-$1/2$ particles.
The overlap of the spin-$1/2$ particles with the $J_\alpha^\Delta$ current can be written as
\begin{equation}
\langle 1/2(p') \vert J^\Delta_\alpha \vert 0 \rangle = \left(A p'_\alpha + B \gamma_\alpha \right) u(p')
\end{equation}
where $u(p')$ is the Dirac spinor describing the spin-$1/2$ particles. Hence, when the gamma matrices are put into the order $\gamma_\alpha\gamma_\mu\gamma_\nu \qslash \pslash' \gamma_5$ in the related correlation function,
the spin-$1/2$ states contribute only to the structures which have $\gamma_\alpha$ at the beginning or those that  are proportional to $p'_\alpha$.  Then, the contributions of the spin-$1/2$ states in the correlation
function are eliminated by ignoring the structures proportional to
$p'_\alpha$ and the structures that contain a $\gamma_\alpha$ at the beginning. By this way, only
the contributions from spin-$3/2$ states are kept (see also Refs.  \cite{Belyaev:1993ss,  Belyaev:1982cd}).

\end{widetext}
As a result, one can decompose  the hadronic representation  of the correlation function in terms of the  various  Lorentz structures entering the calculations:
\begin{align}
 \Pi_{\alpha\mu\nu}^{Had} (p,q) &=
 \Pi_1^{Had}(Q^2)\,  q_\mu g_{\alpha \nu} \qslash \gamma_5   
 + \Pi_2^{Had}(Q^2)\,  p^{\prime}_\mu p^{\prime}_\nu q_\alpha \gamma_5  
 + \Pi_3^{Had}(Q^2)\, q_\mu q_\nu q_\alpha \qslash \gamma_5 
 + \Pi_4^{Had}(Q^2)\,g_{\alpha \mu}\gamma_\nu \qslash \gamma_5\nonumber\\
 &+ \Pi_5^{Had}(Q^2)\,  q_\alpha q_\mu \gamma_\nu \qslash  \gamma_5 
 + \Pi_6^{Had}(Q^2)\, g_{\mu\nu} q_\alpha \gamma_5      
 + \Pi_7^{Had}(Q^2)\,  p^{\prime}_\mu q_\alpha q_\nu \gamma_5  
 + \Pi_8^{Had}(Q^2)\, q_\alpha q_\nu \gamma_\mu \gamma_5 \nonumber\\
&+ \Pi_9^{Had}(Q^2)\, g_{\alpha\mu}q_\nu \gamma_5 +...,
\end{align}
where the invariant functions  $ \Pi_i^{Had}(Q^2) $ are functions of GFFs.

The next step is to calculate the correlation function in quark-gluon language and in terms of the QCD fundamental degrees of freedom. For this purpose, we insert the explicit forms of the currents, which are given  in terms of the corresponding quark fields,  into the  correlation function. These currents are given as:
\begin{align}
\label{intpol}
 J_{\alpha}^{\Delta}(0)=&\frac{1}{\sqrt{3}}\epsilon^{abc}[2(u^{aT}(0) C\gamma_\alpha d^b(0)) u^c(0)+ (u^{aT}(0)C\gamma_\alpha u^b(0))d^c(0)],\nonumber\\
 T_{\mu\nu}^q (x) &= \frac{i}{2}\bigg[\bar{u}^d(x)\overleftrightarrow{D}_\mu(x) \gamma_\nu u^d(x) 
 + \bar{d}^d(x)\overleftrightarrow{D}_\mu(x) \gamma_\nu d^d(x) 
+(\mu \leftrightarrow \nu) \bigg],
\end{align}
where $C$ is the charge conjugation operator; and $a$, $b$, $c$, $d$ are color indices.
The  covariant derivative, $ \overleftrightarrow{D}_\mu(x)  $, is defined as
\begin{align}
 \overleftrightarrow{D}_\mu(x) &=\frac{1}{2} \Big[ \overrightarrow{D}_\mu(x) - \overleftarrow{D}_\mu(x) \Big],
\end{align}
where
\begin{align}
 \overrightarrow{D}_\mu(x) &= \overrightarrow{\partial}_\mu(x)-i\frac{g}{2}\lambda^e A^e_\mu(x), \\
\overleftarrow{D}_\mu(x) &= \overleftarrow{\partial}_\mu(x) +i\frac{g}{2}\lambda^e A^e_\mu(x), 
\end{align}
 with $\lambda^e$ and  $A^e_\mu (x)$ ($ e $ runs from $ 1 $ to $ 8 $)  being the Gell-Mann matrices and external gluon field, respectively.  By using  the explicit forms of the interpolating currents in the correlation function and doing all the possible contractions among the quark fields via the Wick's theorem, we get the QCD side of the correlation function  in terms of the light-quark propagators and  DAs of the nucleon:
 \begin{widetext}
  \begin{eqnarray}\label{corrfunc}
( \Pi_{\alpha \mu\nu}^{QCD} )_{\lambda\eta}(p,q)&=&-\frac{1}{8\sqrt3}\int d^4 x e^{iqx}~\bigg[(C\gamma_\alpha)_{\alpha\beta}( \overleftrightarrow{D}_\mu (x)\gamma_\nu)_{\rho\sigma} + (\mu \leftrightarrow \nu) \bigg] \bigg\{4\epsilon^{abc}\langle \nonumber
0|{u}_{\sigma}^a(0) {u}_{\theta}^b(x) {d}_{\phi}^c(0)|N(p,s) \rangle  \\ \nonumber
& & \times \bigg[2\delta_\alpha^{\eta}\delta_\sigma^{\theta} \delta_\beta^{\phi}S_q(-x)_{\lambda\rho}+
2\delta_\lambda^{\eta} \delta_\sigma^{\theta} \delta_\beta^{\phi}S_q(-x)_{\alpha\rho}\nonumber
+ \delta_\alpha^{\eta} \delta_\sigma^{\theta} \delta_\lambda^{\phi} S_q(-x)_{\beta\rho}+\delta_\beta^{\eta} \delta_\sigma^{\theta} \delta_\phi^{\lambda} S_q(-x)_{\alpha\rho}\bigg]\nonumber \\
& &
- 4\epsilon^{abc}\langle 0|{u}_{\sigma}^a(0) {u}_{\theta}^b(0) {d}_{\phi}^c(x)|N(p,s) \rangle 
\bigg[ 2\delta_\alpha^{\eta}\delta_\lambda^{\theta} \delta_\sigma^{\phi} S_q(-x)_{\beta\rho}
+\delta_\alpha^{\eta}\delta_\beta^{\theta} \delta_\sigma^{\phi} S_q(-x)_{\lambda\rho}\bigg]\bigg\},
\end{eqnarray}
\end{widetext}
where $S_q(x)$ is the light quark propagator defined as
\begin{align}
\label{edmn09}
S_{q}(x) &= 
\frac{1}{2 \pi^2 x^2}\Big( i \frac{{\xslash}}{x^{2}}-\frac{m_{q}}{2 } \Big)
- \frac{\langle  \bar qq \rangle }{12} \Big(1-i\frac{m_{q} \xslash}{4}   \Big)
- \frac{\langle \bar q \sigma.G q \rangle }{192}x^2  \Big(1-i\frac{m_{q} \xslash}{6}   \Big)
-\frac {i g_s }{32 \pi^2 x^2} ~G^{\mu \nu} (x) \bigg[\rlap/{x}
\sigma_{\mu \nu} +  \sigma_{\mu \nu} \rlap/{x}
 \bigg].
\end{align}
We set  $m_q = 0$, and  the terms $ \sim $ $\langle  \bar qq \rangle$ and $\langle \bar q \sigma.G q \rangle$ are omitted   following the  Borel transformation, which is applied to suppress the contributions of the higher states and continuum. Hence, only  the first term in  the light quark propagator gives contribution to  the calculations.
To proceed with the calculation of the correlation function, the matrix element of the local three-quark operator
$4\epsilon^{abc}\langle 0|q_{1\alpha}^a(a_1 x) q_{2\beta}^b(a_2 x) q_{3\gamma}^c(a_3 x)|N(p,s)\rangle$
is needed.
The light-cone distribution amplitudes of the nucleon, which we use in our study to extract the GFFs,  are presented in Ref.~\cite{Braun:2006hz} up to twist six on the basis of QCD conformal partial wave expansion. All parameters inside the DAs are also borrowed from this reference.

By using the  DAs of the  nucleon  and  applying  the  Fourier transformations, the QCD representation of the correlation function is acquired in terms of different Lorentz structures  in the following form:
\begin{align}
 \Pi_{\alpha\mu\nu}^{QCD} (p,q) &=
 \Pi_1^{QCD}(Q^2)\,  q_\mu g_{\alpha \nu} \qslash \gamma_5   
 + \Pi_2^{QCD}(Q^2)\,  p^{\prime}_\mu p^{\prime}_\nu q_\alpha \gamma_5  
 + \Pi_3^{QCD}(Q^2)\, q_\mu q_\nu q_\alpha \qslash \gamma_5 
 + \Pi_4^{QCD}(Q^2)\,g_{\alpha \mu}\gamma_\nu \qslash \gamma_5\nonumber\\
 &+ \Pi_5^{QCD}(Q^2)\,  q_\alpha q_\mu \gamma_\nu \qslash  \gamma_5 
 + \Pi_6^{QCD}(Q^2)\, g_{\mu\nu} q_\alpha \gamma_5      
 + \Pi_7^{QCD}(Q^2)\,  p^{\prime} _\mu q_\alpha q_\nu \gamma_5  
 + \Pi_8^{QCD}(Q^2)\, q_\alpha q_\nu \gamma_\mu \gamma_5 \nonumber\\
&+ \Pi_9^{QCD}(Q^2)\, g_{\alpha\mu}q_\nu \gamma_5 +... ~,
\end{align}
where $ \Pi_{1,2,...,9}^{QCD}(Q^2) $ are invariant functions corresponding to the coefficients of the selected structures.

The   LCSR for the $N \rightarrow \Delta $ transition GFFs are obtained  by matching  the coefficients of 
the chosen Lorentz structures from both the physical  and QCD representations  of the correlation function.  To abolish the contributions  coming from the higher states and continuum, Borel transformation as well as  continuum subtraction supplied by the quark-hadron duality assumption are applied.
We should stress that, we use the Lorentz structures $q_\mu g_{\alpha \nu} \qslash \gamma_5 $, $p^{\prime}_\mu p^{\prime}_\nu q_\alpha \gamma_5$, 
$q_\mu q_\nu q_\alpha \qslash \gamma_5$,  $g_{\alpha \mu}\gamma_\nu \qslash \gamma_5$, $q_\alpha q_\mu \gamma_\nu \qslash  \gamma_5$,  $g_{\mu\nu} q_\alpha \gamma_5$, $ p^{\prime}_\mu q_\alpha q_\nu \gamma_5$, $q_\alpha q_\nu \gamma_\mu \gamma_5$ and $g_{\alpha\mu}q_\nu \gamma_5 $ to find the LCSR for  the $N \rightarrow \Delta $ transition GFFs, $F_1^{N\Delta}(Q^2)$, $F_2^{N\Delta}(Q^2)  $, $F_3^{N\Delta}(Q^2) $, $F_4^{N\Delta}(Q^2) $, $F_5^{N\Delta}(Q^2) $,  $\bar C_1^{N\Delta}  (Q^2) $, $\bar C_2^{N\Delta}  (Q^2) $, $\bar C_3^{N\Delta}  (Q^2) $, and $\bar C_4^{N\Delta} (Q^2) $, respectively.
 Hence, 
\begin{align}
&
F_1^{N\Delta}(Q^2) \frac{\, \lambda_\Delta}{\, (m_\Delta^2-{p^{\prime}}^2)}  = -2\,\rho_1^{QCD}(v,y,x_1,x_2,x_3),~~~~ ~~~~ ~~~~
F_2^{N\Delta}(Q^2) \frac{\lambda_\Delta (m_\Delta+ m_N) }{(m_\Delta^2-{p^{\prime}}^2)}= \bar m^2 \rho_2^{QCD}(v,y,x_1,x_2,x_3),\nonumber\\  
& F_3^{N\Delta}(Q^2) \frac{\lambda_\Delta}{(m_\Delta^2-{p^{\prime}}^2)}  =- 2\,\bar m^2 \rho_3^{QCD}(v,y,x_1,x_2,x_3),
 ~~~~~~~~   
 F_4^{N\Delta}(Q^2) \frac{\lambda_\Delta}{(m_\Delta^2-{p^{\prime}}^2)}  =- \frac{3}{\bar m}\rho_4^{QCD}(v,y,x_1,x_2,x_3), \nonumber\\ 
& F_5^{N\Delta}(Q^2) \frac{\lambda_\Delta }{(m_\Delta^2-{p^{\prime}}^2)}  = 2\,\bar m \, \rho_5^{QCD}(v,y,x_1,x_2,x_3), ~~~~~~ ~~~~~   
 \bar C_1^{N\Delta} (Q^2) \frac{\lambda_\Delta (m_\Delta+ m_N)}{(m_\Delta^2-{p^{\prime}}^2)} = \rho_6^{QCD}(v,y,x_1,x_2,x_3),\nonumber\\   
&\bar C_2^{N\Delta} (Q^2) \frac{(m_\Delta+ m_N)\,\lambda_\Delta}{\,(m_\Delta^2-{p^{\prime}}^2)}  =  \frac{\bar m^2}{2} \, \rho_7^{QCD}(v,y,x_1,x_2,x_3), ~~~~ ~~ 
  \bar C_3^{N\Delta} (Q^2) \frac{\lambda_\Delta (m_\Delta+ m_N)}{(m_\Delta^2-{p^{\prime}}^2)} = \bar m \, \rho_8^{QCD}(v,y,x_1,x_2,x_3), \nonumber\\ 
&  \bar C_4^{N\Delta} (Q^2) \frac{\lambda_\Delta}{(m_N+ m_\Delta)\,(m_\Delta^2-{p^{\prime}}^2)}  =   \rho_9^{QCD}(v,y,x_1,x_2,x_3).\label{C4FF}
\end{align}
 The explicit expressions of the $\rho_i^{QCD}(v,y,x_1,x_2,x_3)$ functions are given in the Appendix. 
%

\section{Numerical Results}\label{secIII}

 The sum rules obtained for the GFFs in the previous section contain different variables: Hadronic and  QCD input parameters, input parameters inside the DAs of the nucleon, some auxiliary or helping parameters and transferred momentum square $ Q^2 $. The main purpose in this section is to discuss the $ Q^2 $ behavior of the GFFs. 
 To this end, the input parameters of the quarks, nucleon and $\Delta  $ baryon entering the sum rules are selected as  $m_u =m_d =0$, $m_N=0.94$~GeV,  $m_\Delta = 1.23$~GeV and $\lambda_\Delta = 0.038$~GeV$^3$~\cite{Aliev:2007pi,Aliev:2004ju}.  Various input parameters inside the DAs of the nucleon in two sets are borrowed from Ref.~\cite{Braun:2006hz} and presented  in table \ref{parameter_table}.
 \begin{table}[htp]
\renewcommand{\arraystretch}{1.3}
\addtolength{\tabcolsep}{10pt}
$$
\begin{array}{|l|c|c|}
\hline \hline
  \mbox{Parameters}         &  \mbox{Set-I}                       &  \mbox{Set-II}               \\  \hline \hline
 ~~~~~f_N       & ( 5.0 \pm 0.5)\times 10^{-3}~\mbox{GeV}^2  &  ( 5.0 \pm 0.5)\times 10^{-3}~\mbox{GeV}^2                    \\
 ~~~~~\lambda_1 & (-2.7 \pm 0.9)\times 10^{-2}~\mbox{GeV}^2  &  (-2.7 \pm 0.9)\times 10^{-2}~\mbox{GeV}^2                   \\
 ~~~~~\lambda_2 & ( 5.4 \pm 1.9)\times 10^{-2}~\mbox{GeV}^2  &  (5.4 \pm 1.9)\times 10^{-2}~\mbox{GeV}^2                    \\ \hline \hline
 ~~~~~A_1^u     & 0.38 \pm 0.15                       & 0                         \\
 ~~~~~V_1^d     & 0.23 \pm 0.03                       & 1/3                          \\
 ~~~~~f_1^d     & 0.40 \pm 0.05                       & 1/3                                            \\
~~~~~ f_2^d     & 0.22 \pm 0.05                       & 4/15                                           \\
 ~~~~~f_1^u     & 0.07 \pm 0.05                       & 1/10                                          \\ \hline \hline
\end{array}
$$
\caption{The numerical values of the main input parameters entering the expressions of the nucleon's DAs. The upper panel shows the dimensionfull parameters the nucleon. In the lower panel we list the values of the five dimensionless parameters that determine the shape of the DAs.}
\renewcommand{\arraystretch}{1}
\addtolength{\arraycolsep}{-1.0pt}
\label{parameter_table}
\end{table}
 
\begin{figure}[htp]
\centering
 \subfloat[]{\includegraphics[width=0.33\textwidth]{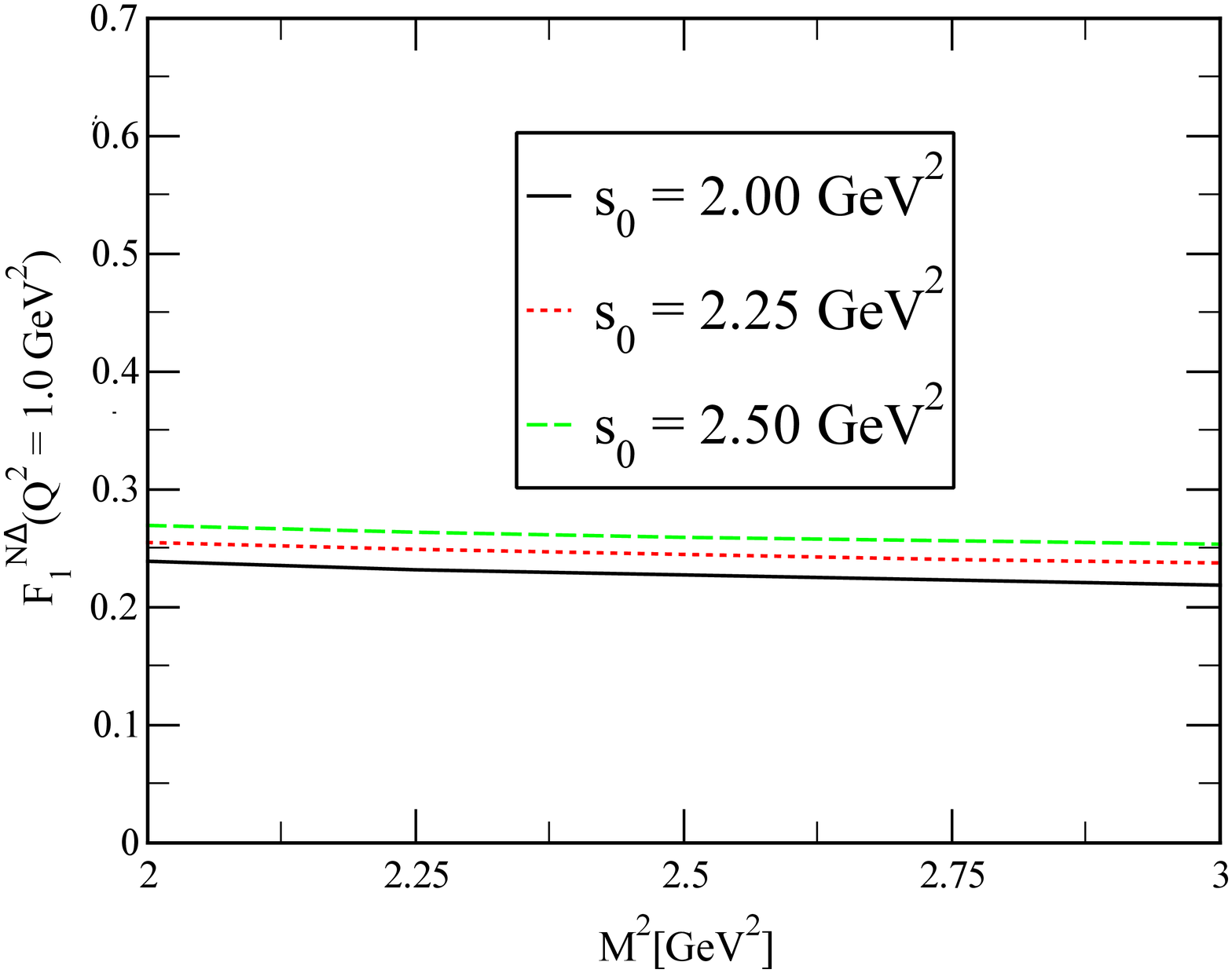}}~~~~~~~~~~
 \subfloat[]{\includegraphics[width=0.33\textwidth]{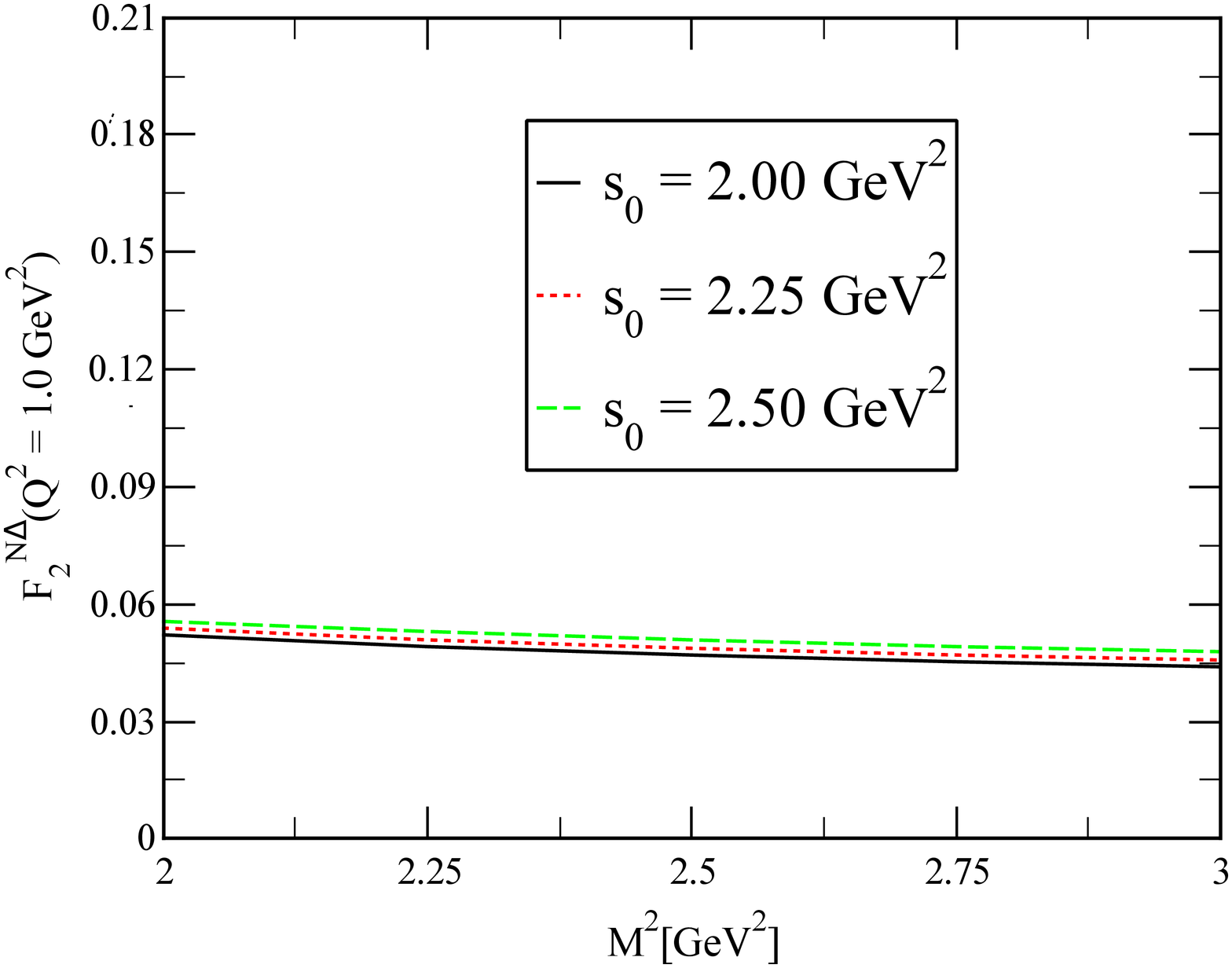}}\\
 \subfloat[]{\includegraphics[width=0.33\textwidth]{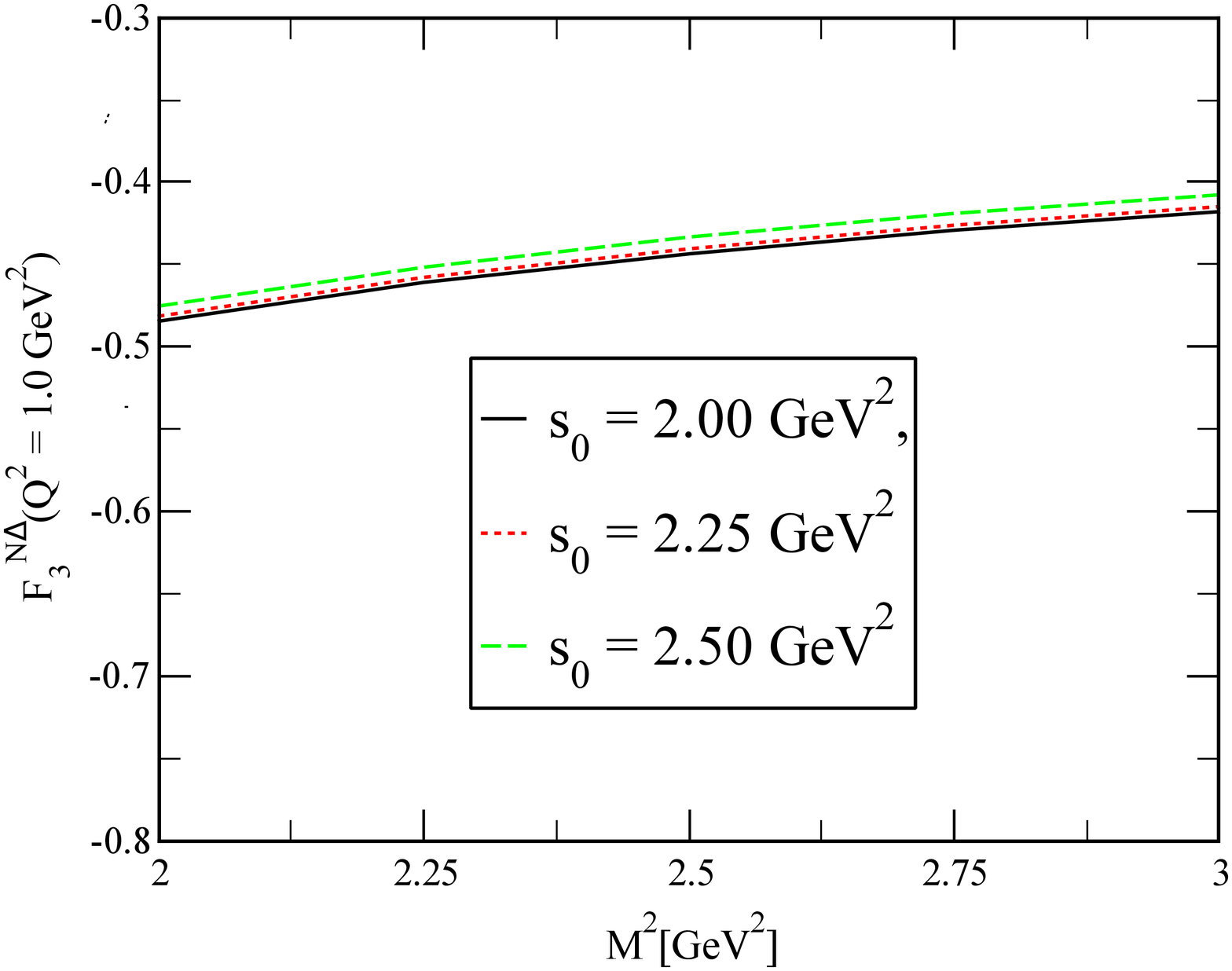}}~~~~~~~~~~
 \subfloat[]{\includegraphics[width=0.33\textwidth]{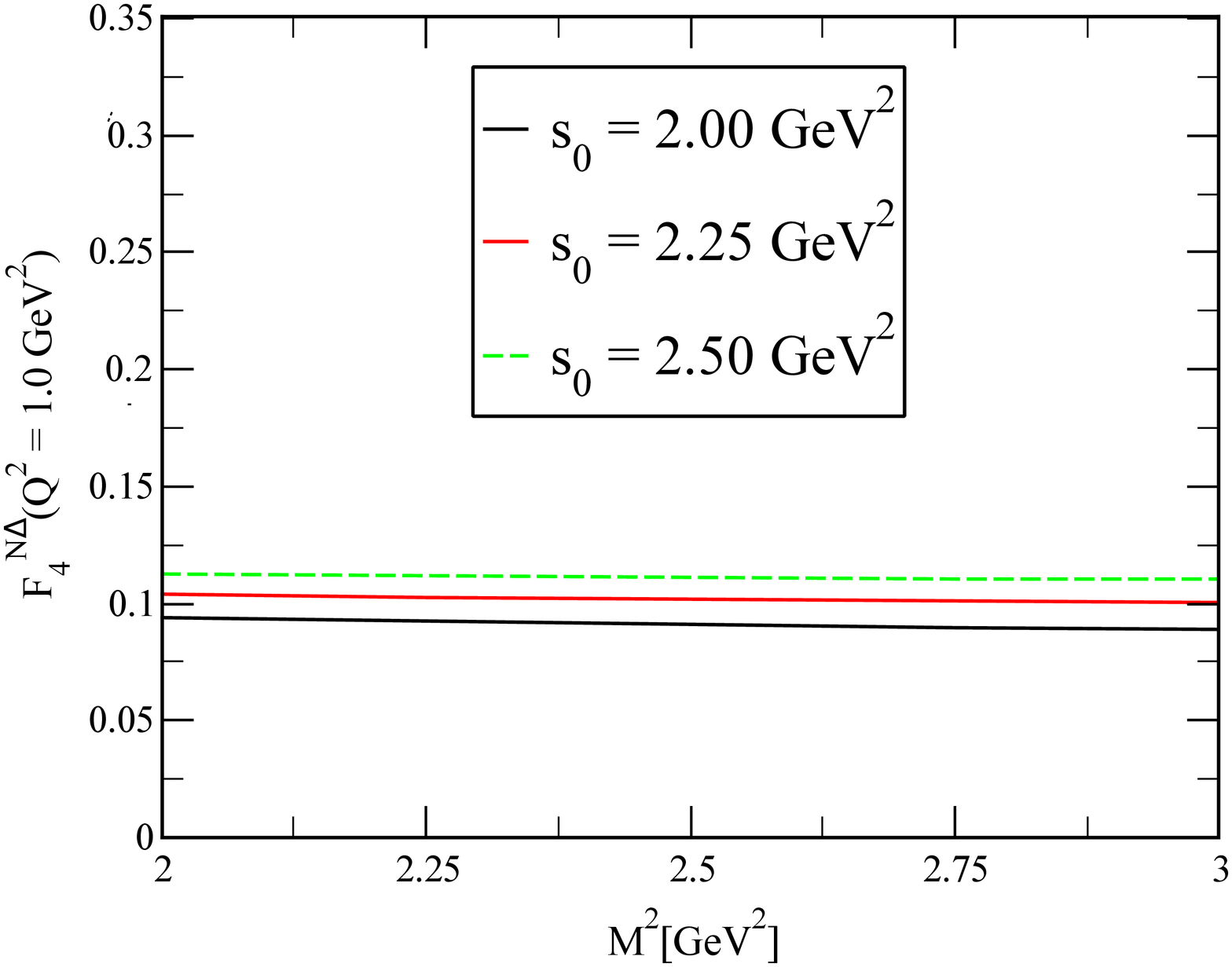}}\\
 \subfloat[]{\includegraphics[width=0.33\textwidth]{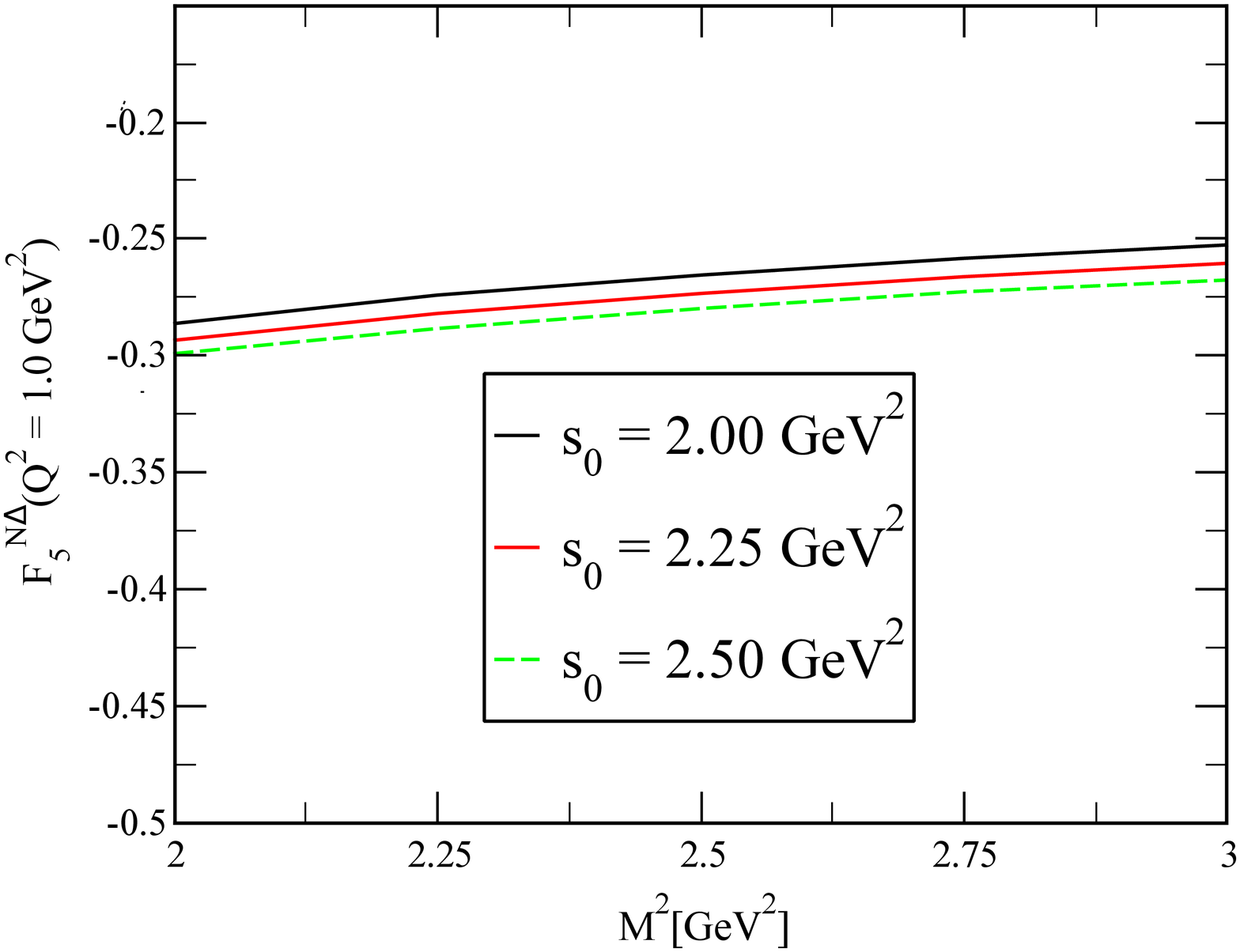}}~~~~~~~~~~
 \subfloat[]{\includegraphics[width=0.33\textwidth]{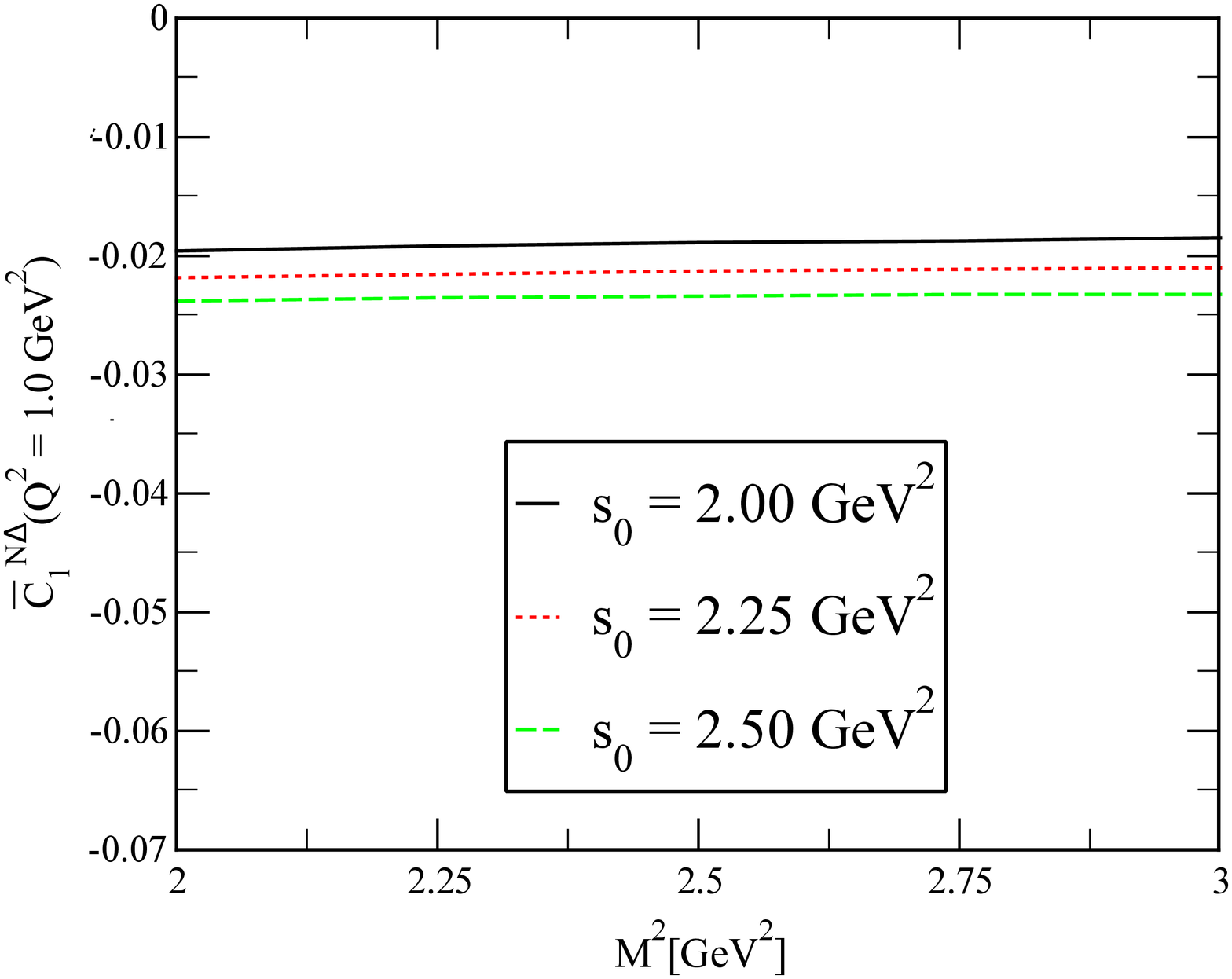}}\\
 \subfloat[]{\includegraphics[width=0.33\textwidth]{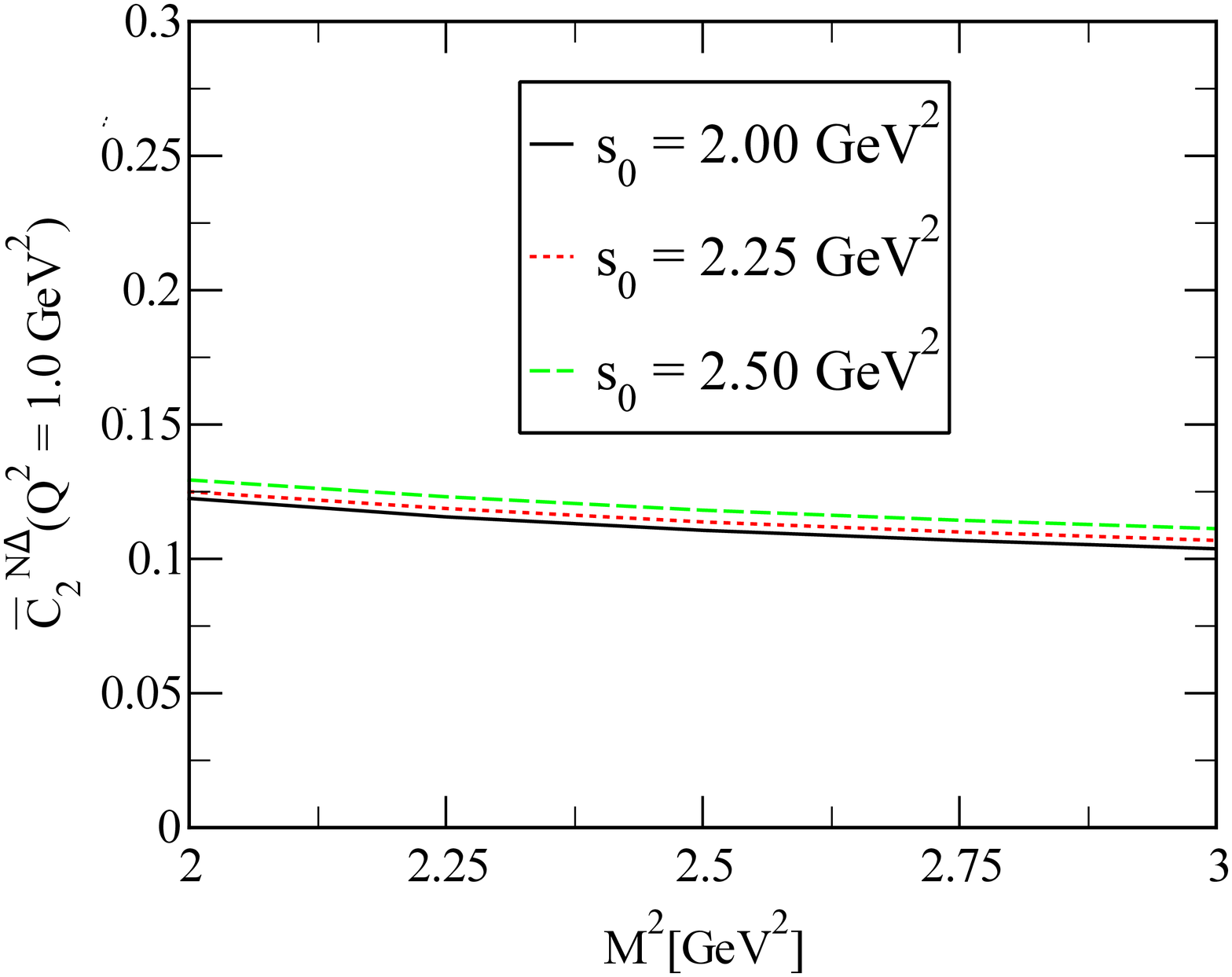}}~~~~~~~~~~
 \subfloat[]{\includegraphics[width=0.33\textwidth]{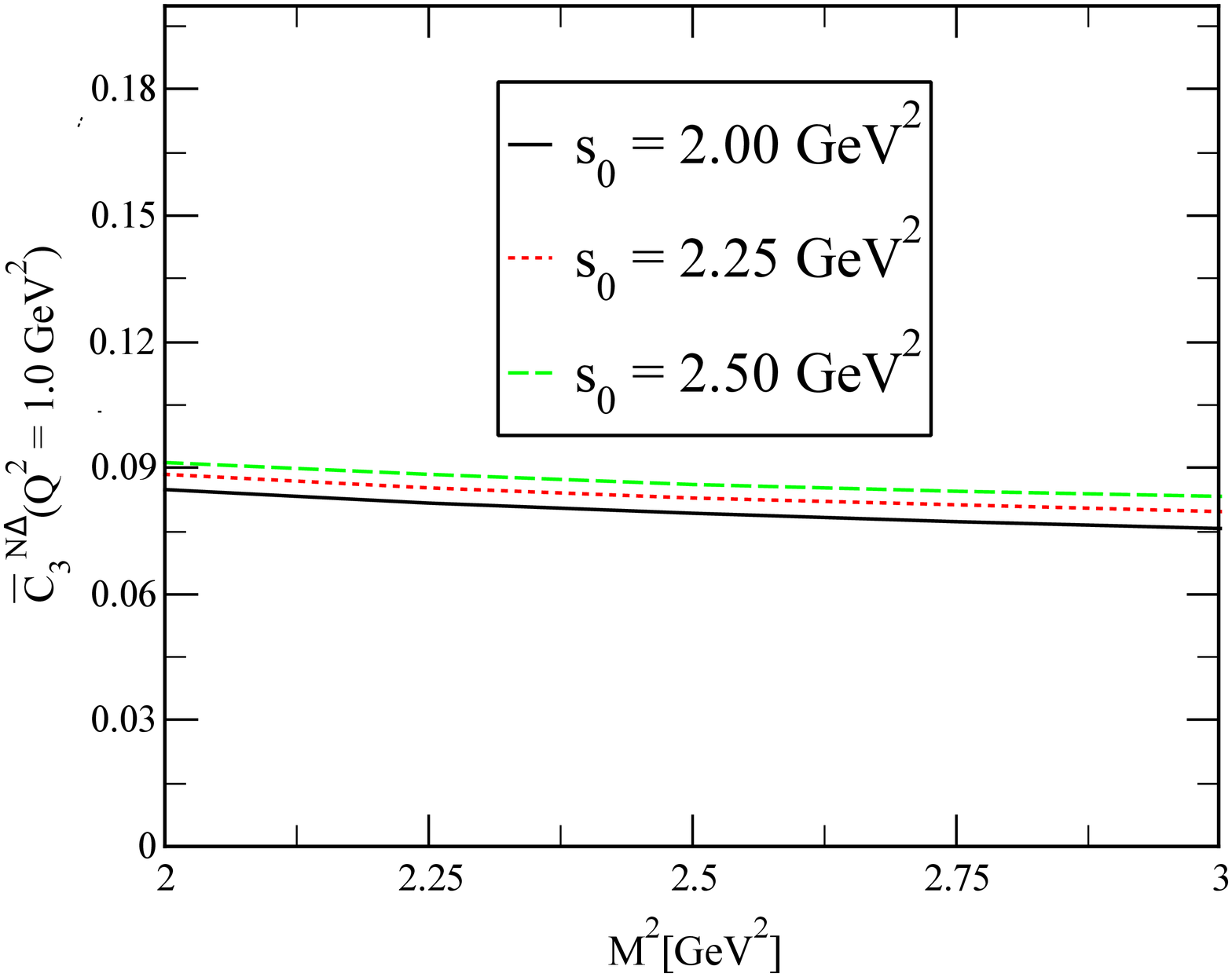}}
 \caption{The dependence of the $N \rightarrow \Delta$ transition GFFs on $M^2$ at  $Q^2 = 1.0$~GeV$^2$ and three fixed values of the  $s_0$ and   set-I parameters.}
 \label{Msqfigs1}
  \end{figure}

\begin{figure}[htp]
\centering
 \subfloat[]{\includegraphics[width=0.33\textwidth]{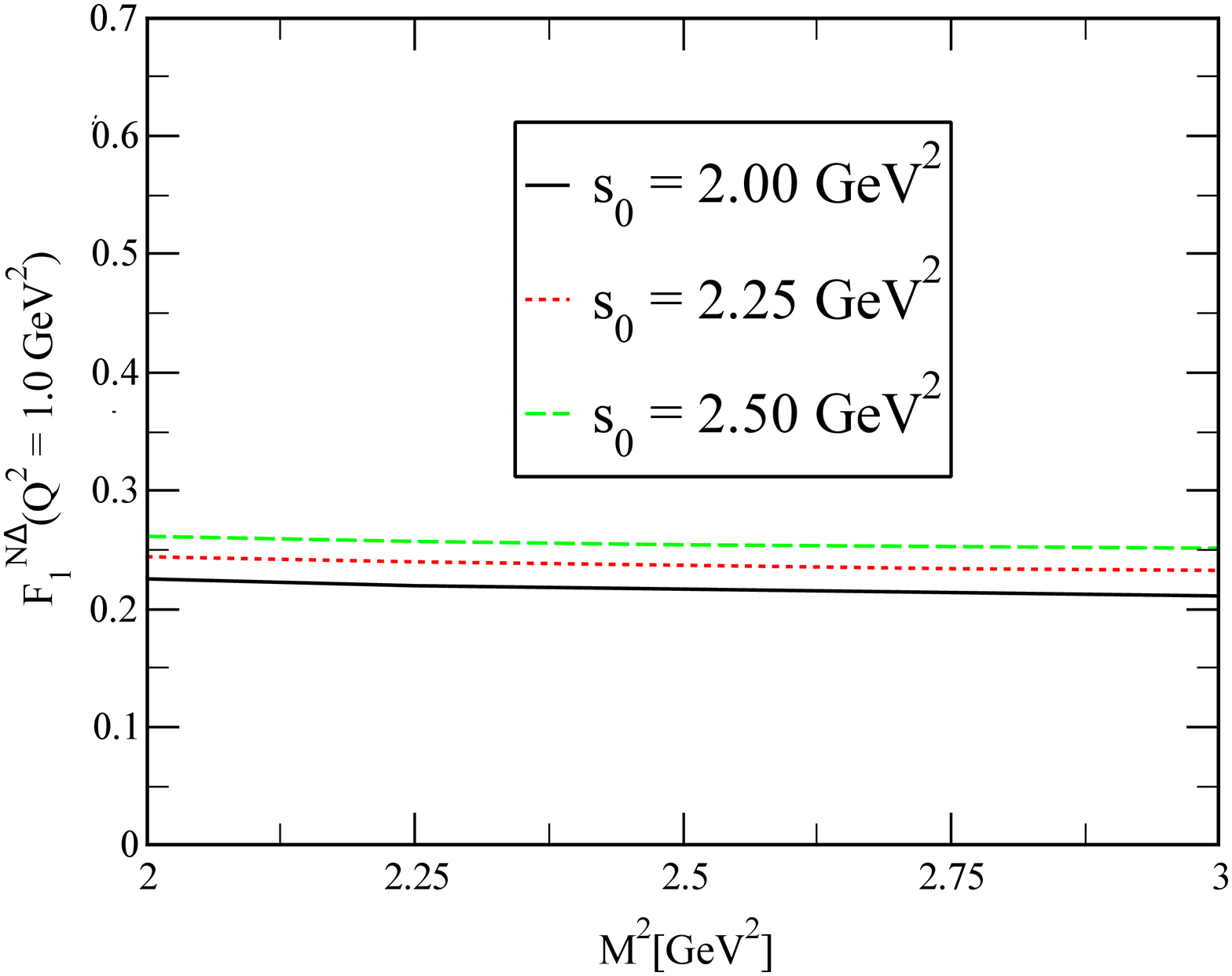}}~~~~~~~~~~
 \subfloat[]{\includegraphics[width=0.33\textwidth]{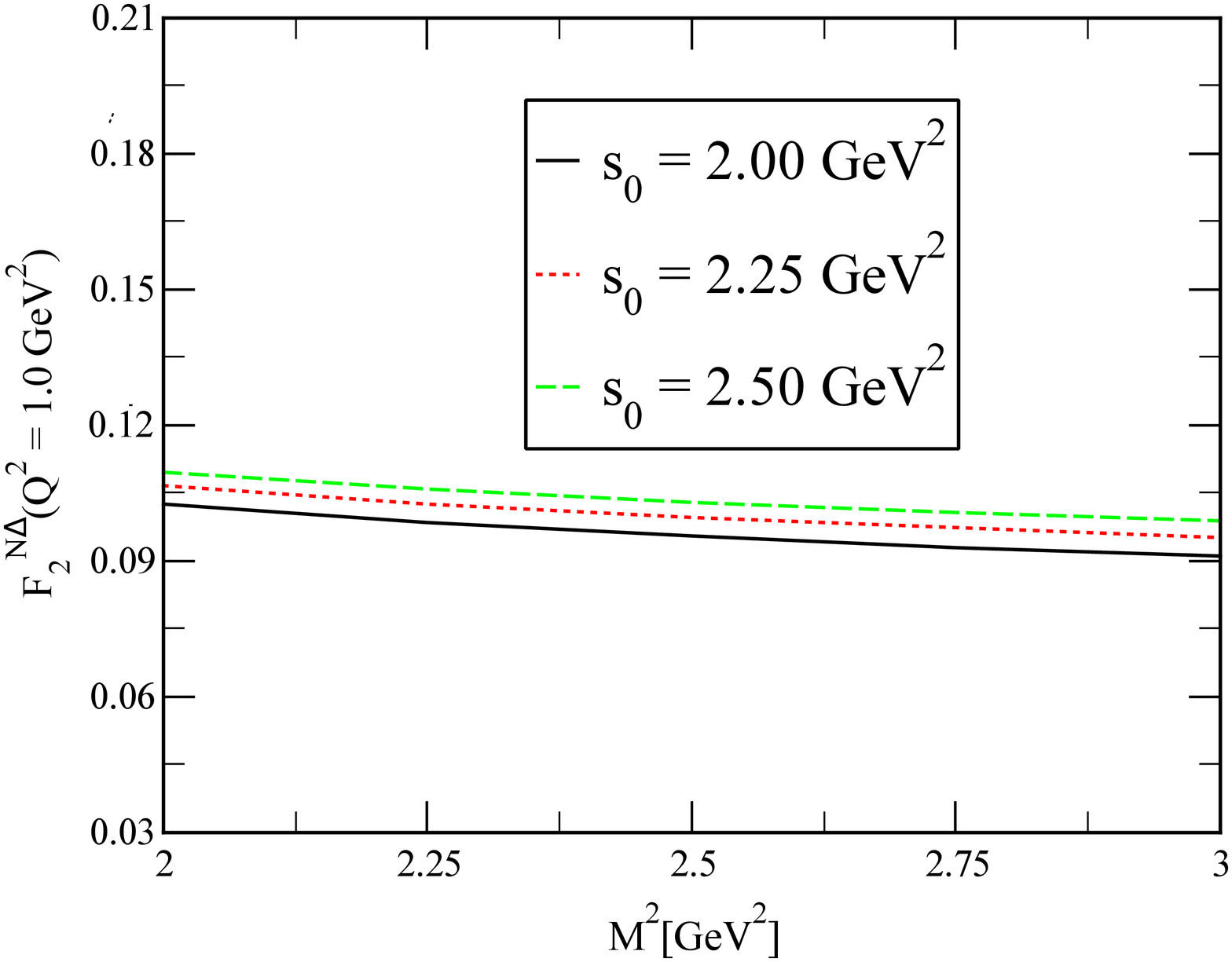}}\\
 \subfloat[]{\includegraphics[width=0.33\textwidth]{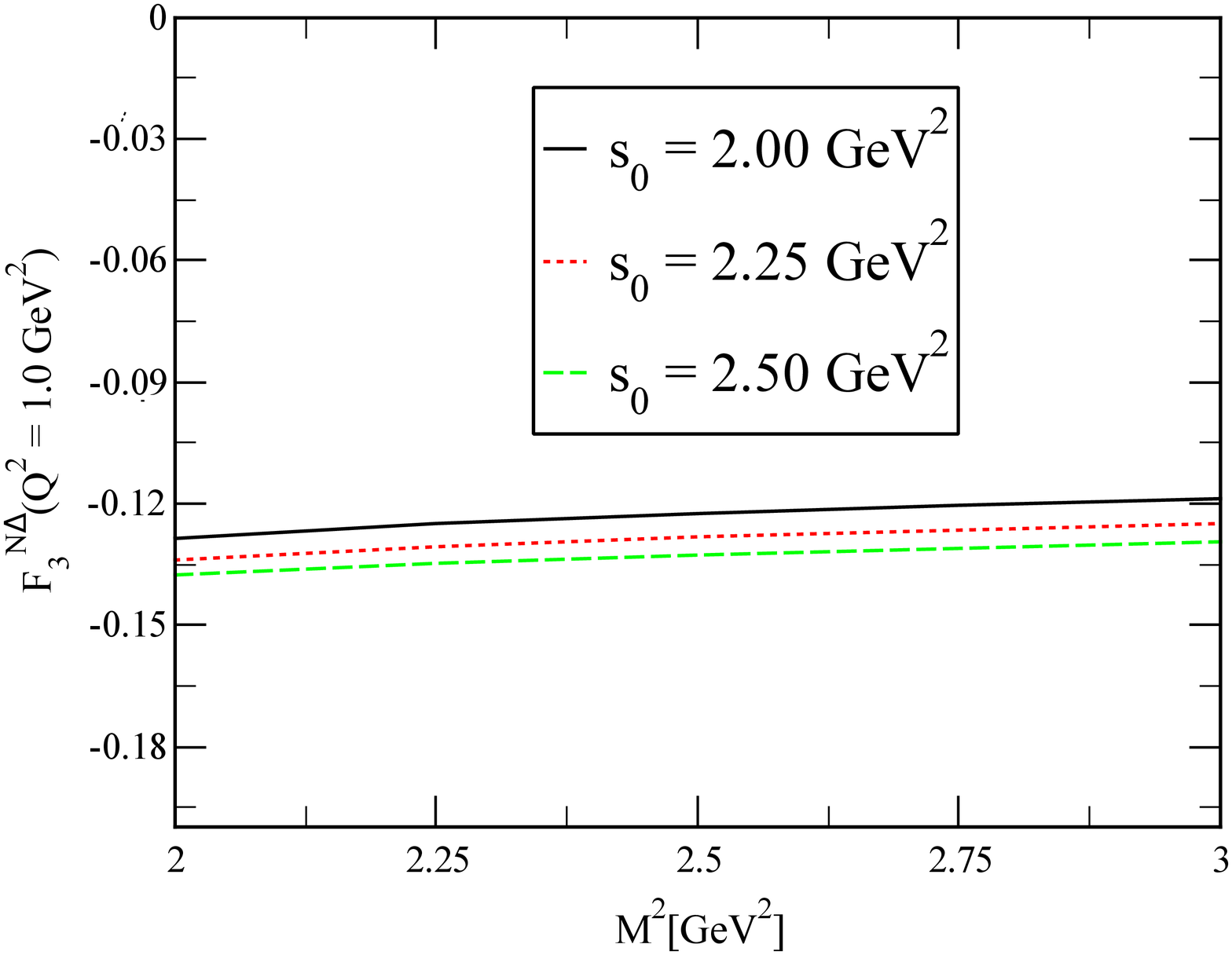}}~~~~~~~~~~
 \subfloat[]{\includegraphics[width=0.33\textwidth]{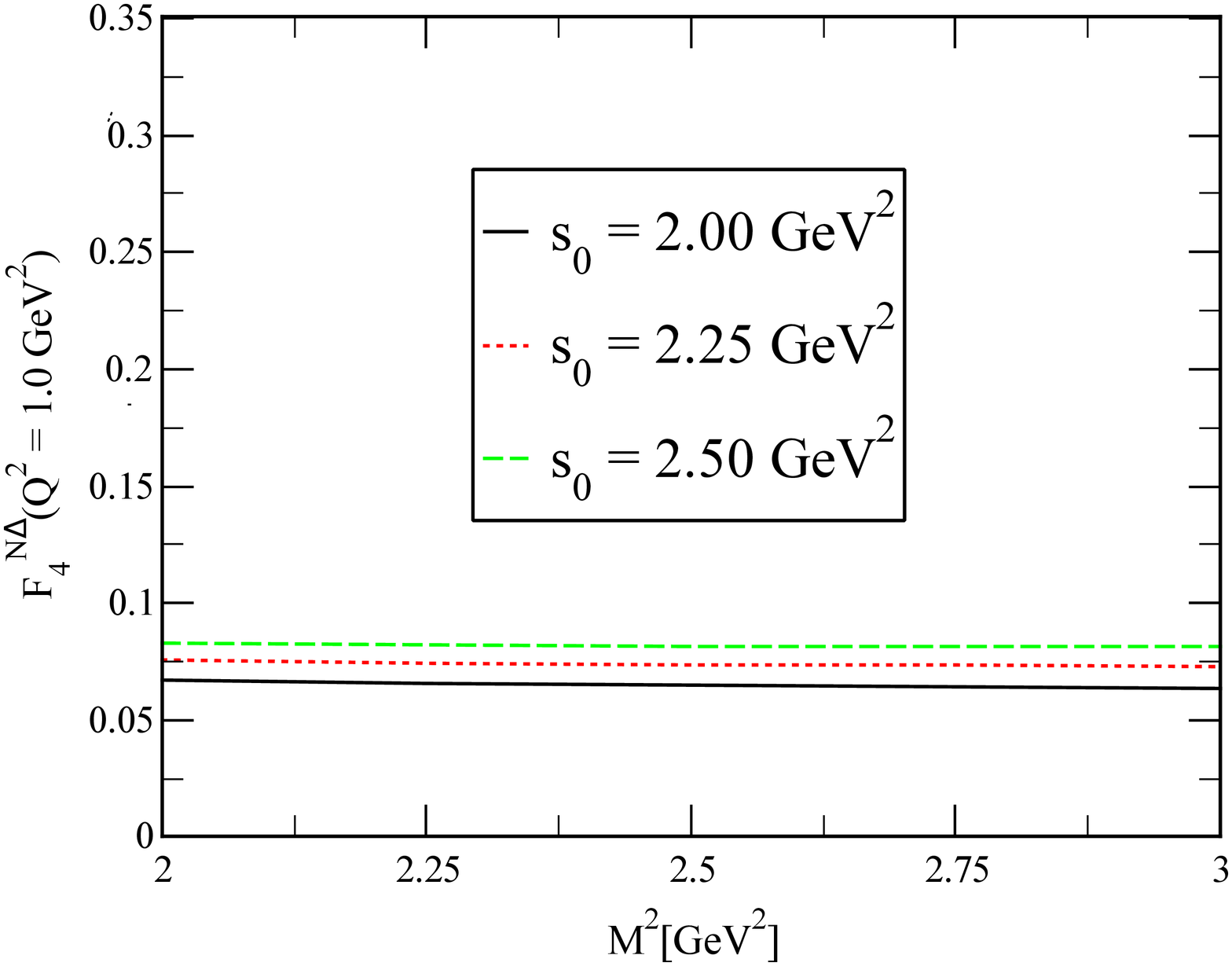}}\\
 \subfloat[]{\includegraphics[width=0.33\textwidth]{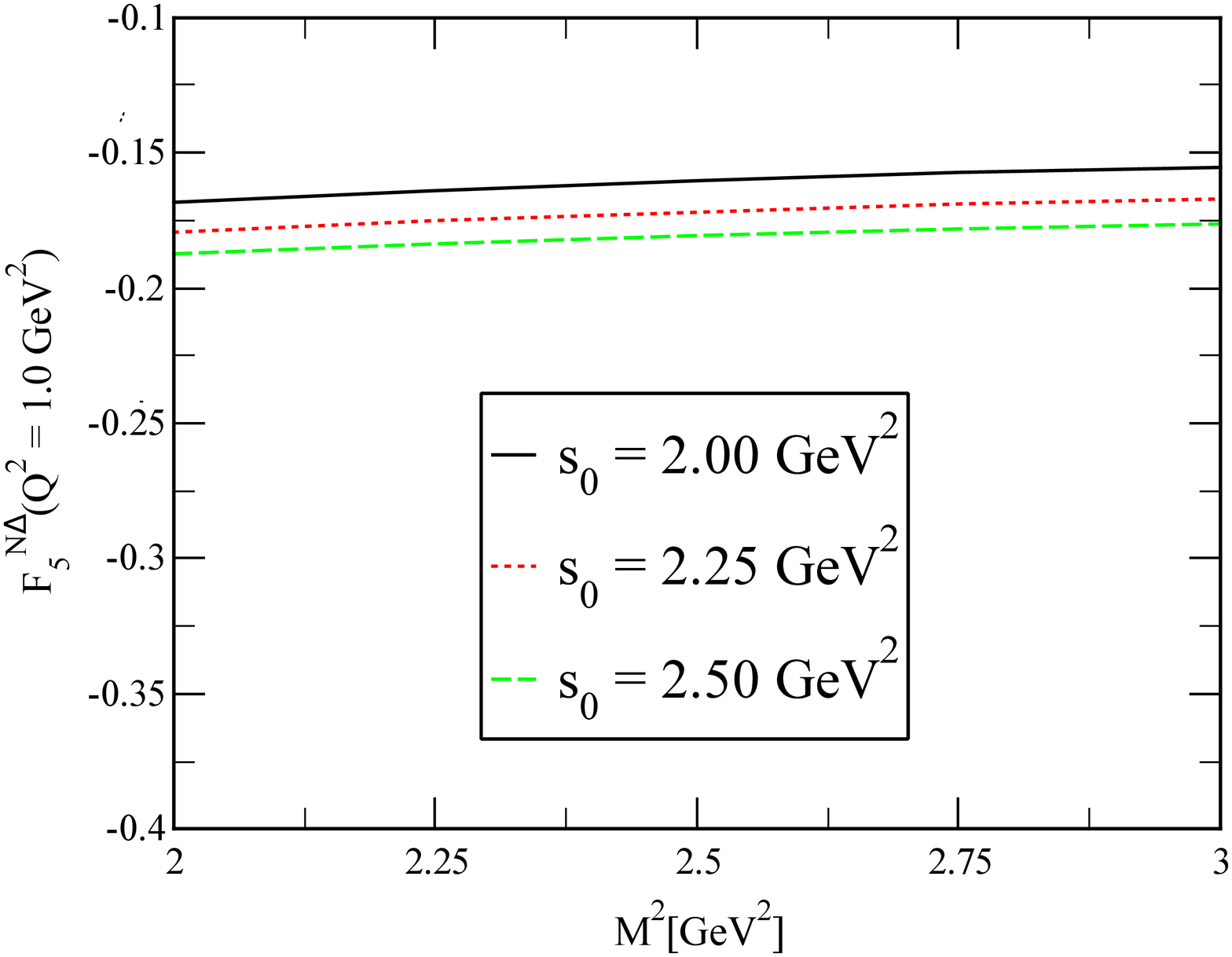}}~~~~~~~~~~
 \subfloat[]{\includegraphics[width=0.33\textwidth]{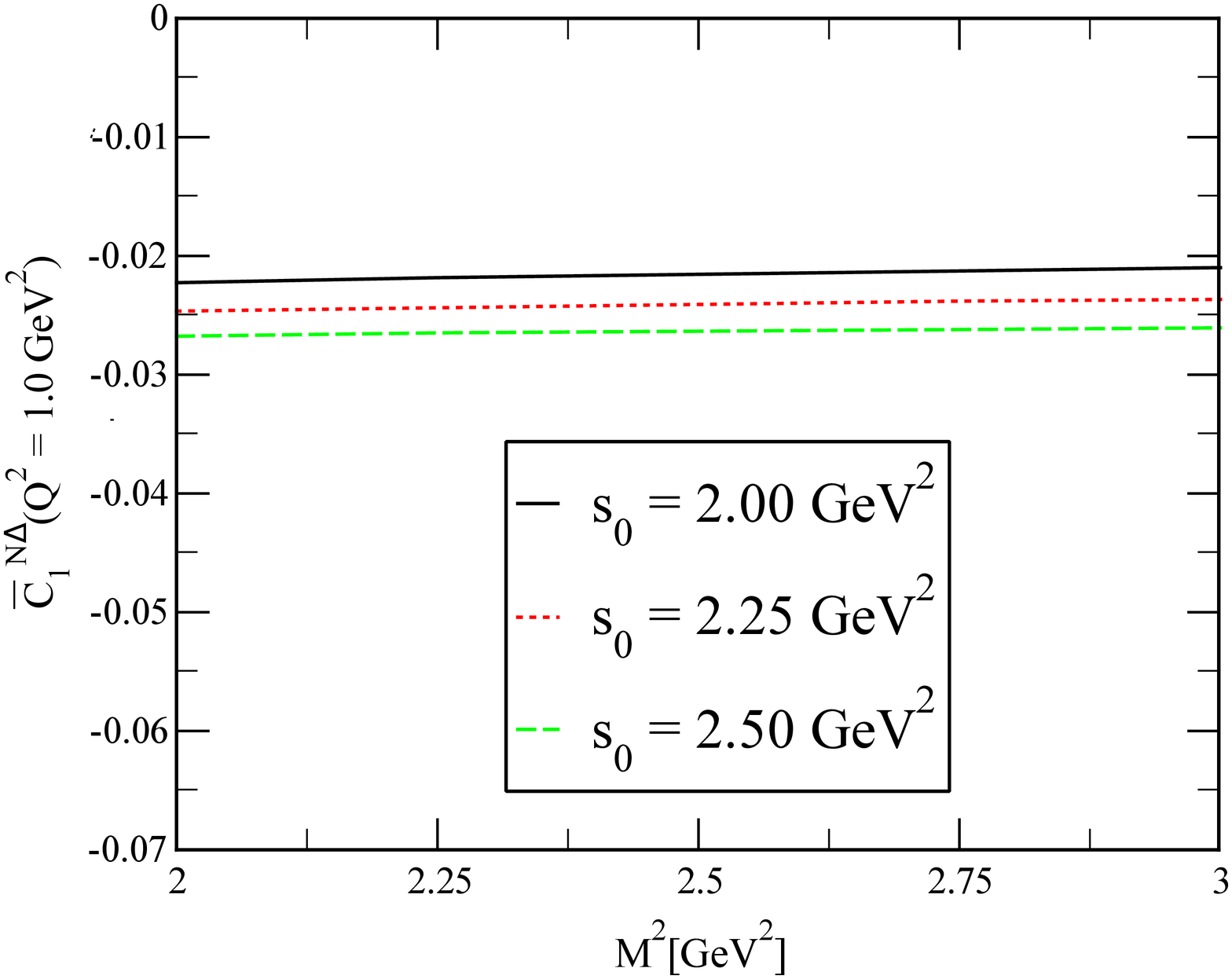}}\\
 \subfloat[]{\includegraphics[width=0.33\textwidth]{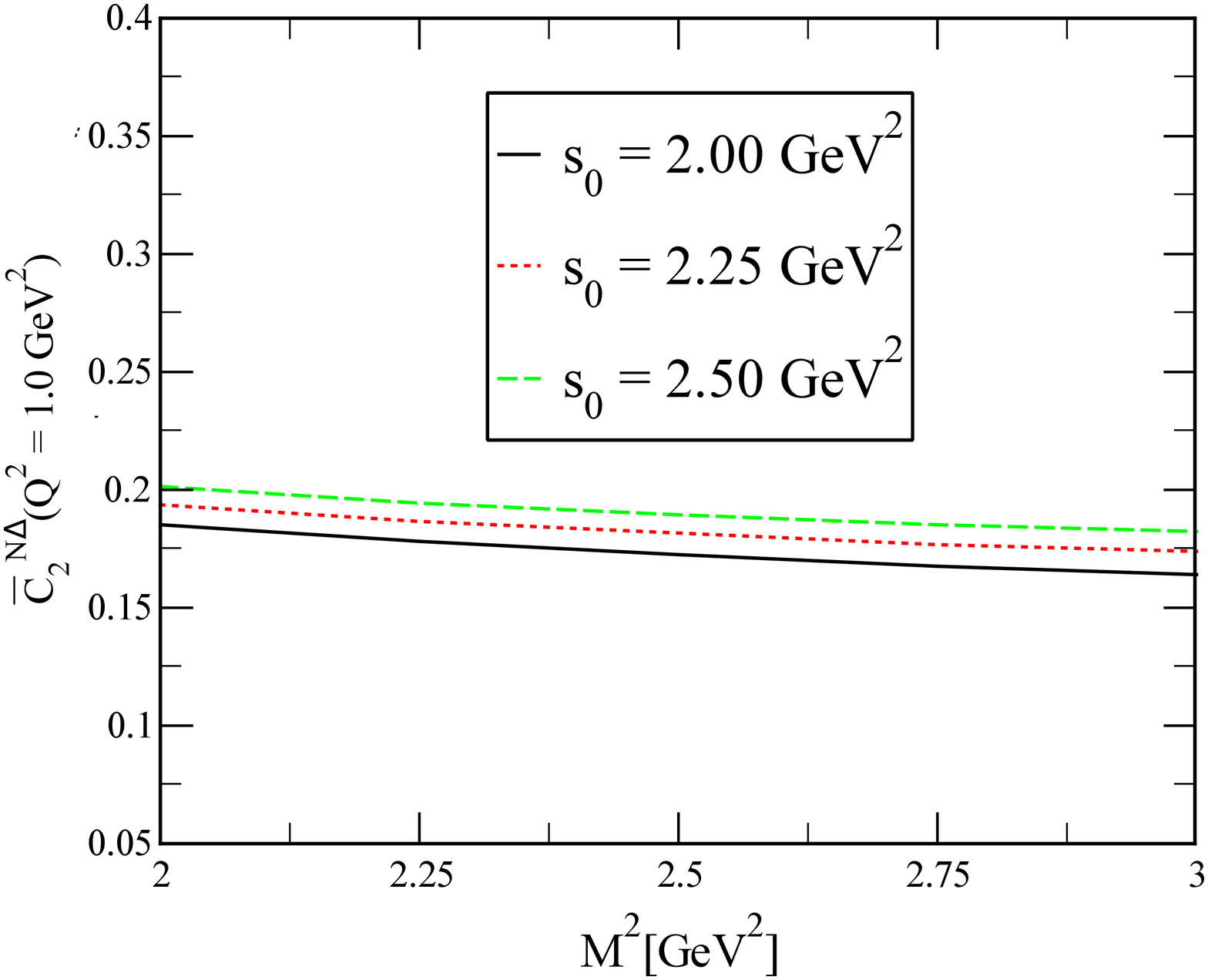}}~~~~~~~~~~
 \subfloat[]{\includegraphics[width=0.33\textwidth]{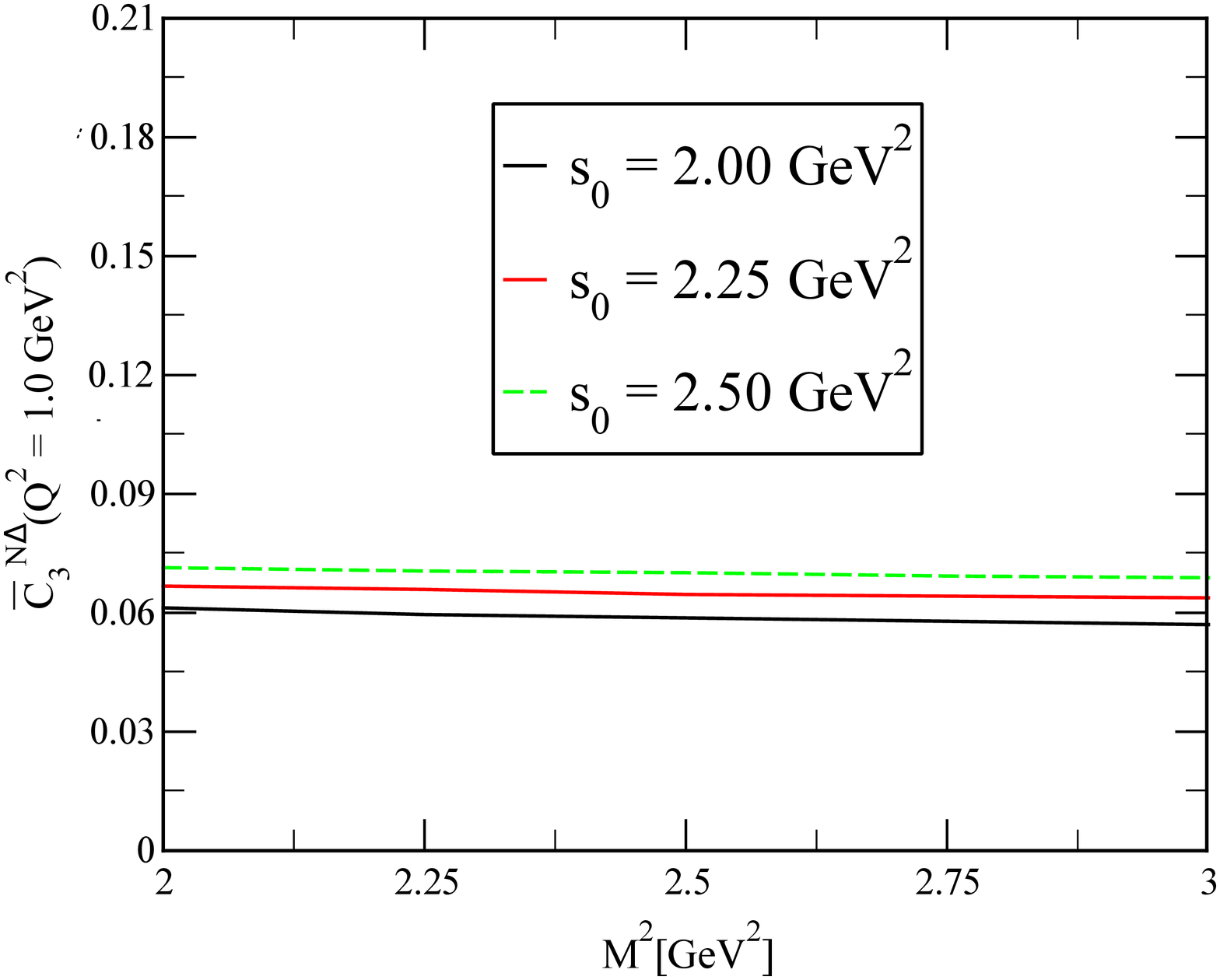}}
 \caption{The dependence of the $N \rightarrow \Delta$ transition GFFs on $M^2$ at  $Q^2 = 1.0$~GeV$^2$ and three fixed values of the  $s_0$ and set-II parameters.}
 \label{Msqfigs2}
  \end{figure}

 The next step is to fix the auxiliary parameters: The Borel parameter $ M^2 $ and the continuum threshold $ s_0 $. To this end, we refer to the standard criteria of the sum rule method: The pole dominance over the contributions of the higher states and continuum as well as the higher the twist of the nucleon DAs, the lower its contribution. These criteria are satisfied in the regions that the GFFs, as physical quantities, depend relatively weakly on the auxiliary parameters and show relatively good stability in their working windows. We acquire the following working windows for the  $ s_0 $ and $ M^2 $ from the analyses:
 \begin{align}
  2.0 ~\mbox{GeV}^2 \leq s_0 \leq 2.50 ~\mbox{GeV}^2, \nonumber\\
  2.0 ~\mbox{GeV}^2 \leq M^2 \leq 3.0 ~\mbox{GeV}^2. \nonumber
 \end{align}
Figures \ref{Msqfigs1} and   \ref{Msqfigs2} present the dependence of the GFFs for $ N-\Delta $ transition on the auxiliary parameters  at $Q^2 = 1.0$ GeV$^{2}$ for 
two sets of the nucleon DAs parameters. These figures demonstrate good stability of the GFFs with respect to the changes of the auxiliary parameters in their working regions: The residual dependencies appear as the uncertainties in the results.
 
 \begin{table}[htp]
\begin{tabular}{ |l|c|c|c|c|c|l|}
\hline\hline
&\multicolumn{3}{|c|}{Results of set-I} &\multicolumn{3}{|c|}{Results of set-II}\\
\hline\hline

Form Factors & ~~~~${f}(0)$(GeV$^{-2}$)]~~~~  & ~~${\cal M}$(GeV)~~ & ~~$\alpha$~~ 
& ~~~~${ f}(0)$(GeV$^{-2}$)~~~~  &~~ ${\cal M}$(GeV)~~ & ~~~~$\alpha$~~ \\ 
\hline\hline
        ~~$F_1^{N\Delta}$($Q^2$)       &~~$0.80 \pm 0.08$    &$ 1.17 \pm 0.12$ &$2.1-2.3$ 
        &$~~1.10 \pm 0.20$    &$ 1.18 \pm 0.10$ &$2.8-3.0$\\
        ~~$F_2^{N\Delta}$($Q^2$)       &~~$0.20 \pm 0.03$    &$1.11 \pm 0.05$  &$2.4-2.6$  
        &$~~0.31 \pm 0.04$    &$1.26 \pm 0.11$  &$2.2-2.4$\\
        ~~$F_3^{N\Delta}$($Q^2$)       &$-1.57 \pm 0.26$   &$1.22 \pm 0.10$  &$2.3-2.5$   
        &$-0.97 \pm 0.12$   &$1.14 \pm 0.11$  &$3.4-3.6$  \\
        ~~$F_4^{N\Delta}$($Q^2$)       &~~$0.38 \pm 0.06$    &$1.17 \pm 0.10$  &$2.4-2.6$ 
        &$~~0.28 \pm 0.03$    &$1.20 \pm 0.11$  &$2.4-2.6$\\
        ~~$F_5^{N\Delta}$($Q^2$)       &$-0.51 \pm 0.05$   & $1.18 \pm 0.09$ &$2.2-2.4$ 
        &$-0.71 \pm 0.11$   & $1.10 \pm 0.13$ &$2.2-2.4$\\
       ~~$\bar C_1^{N\Delta}$($Q^2$)  &~$-0.073 \pm 0.003$ &$ 1.24 \pm 0.10$ &$2.4-2.6$  
        &$~-0.083 \pm 0.003$ &$ 1.24 \pm 0.11$ &$2.4-2.6$ \\
        ~~$\bar C_2^{N\Delta}$($Q^2$)  &~~$0.35 \pm 0.03$    &$1.29 \pm 0.05$  &$2.3-2.5$  
        &$~~0.63 \pm 0.07$    &$1.16 \pm 0.10$  &$2.1-2.3$ \\
        ~~$\bar C_3^{N\Delta}$($Q^2$)  &~~$0.20 \pm 0.02$    &$1.29 \pm 0.03$  &$1.8-2.0$ 
        &$~~0.28 \pm 0.03$    &$1.19 \pm 0.10$  &$2.7-2.9$\\
        ~~$\bar C_4^{N\Delta}$($Q^2$)  &~~$-$  &$-$ &$-$  &$~~-$  &$-$ &~~~~$-$ \\
\hline \hline
\end{tabular}
\caption{Numerical values of the fitting parameters for the transition  GFFs of $N-\Delta$.
}
\label{fit_table}
\end{table}

Now, we proceed to discuss the $Q^2 $-behavior of the GFFs. The light-cone QCD sum rules give reliable results for $Q^2 \geq 1$ ~GeV$^2$. 
The mass corrections of the DAs $\sim m^2_{N}/Q^2$ turn out to be  very large for $Q^2 < 1.0$ GeV$^2$, in other words, the light-cone  sum rules become unreliable.  Hence, for the GFFs, we expect the light-cone QCD sum rules to work effectively in the $1.0 $ GeV$^2 \leq Q^2  \leq 10.0$ GeV$^2$ region.
Therefore to find the values of the GFFs at static  limit, $Q^2 =0$, we need to extrapolate the results to small values of $Q^2 $. The following $ \alpha $-pole fit functions do this job well and produce all our sum rules results  for $Q^2 \geq 1$~GeV$^2$:
  \begin{align}
 F_i^{N\Delta}(Q^2)[\bar C_i^{N\Delta}(Q^2)]=  f(0)\bigg[1+ \frac{Q^2}{{\cal M}^2}\bigg]^{-\alpha},
\end{align} 
 where the fit parameters including the values of GFFs at static limit, $ f(0) $, are given in table \ref{fit_table}. The uncertainties in the presented values belong to the errors of all input parameters and those related to the determination of working intervals for the auxiliary parameters. As we previously mentioned,  the quark part of the EMT current  alone is not conserved and  matrix element of the EMT current is parameterized in terms of five independent conserved and four independent non-conserved form factors. The $ C_i^{N\Delta}(Q^2) $ form factors depict the orders of breaking the conservation of quark part of the EMT current. Our analyses do not give reliable results for $ C_4^{N\Delta}(Q^2) $ for both the sets of nucleon DAs parameters.
To our best knowledge, this is the first direct study using an analytic method in the literature dedicated to the calculation of the $N \rightarrow \Delta$ transition GFFs. These GFFs can also be obtained from the transition  GPDs. However, the connection between the transition GPDs and transition energy-momentum tensor form factors is unknown  \cite{Kim:2022bwn}. One should first study these connections like those in the case of nucleon discussed previously. The $N \rightarrow \Delta$ transition GPDs were studied in the large $N_c  $ limit in Ref. \cite{Goeke:2001tz}. It will be possible to extract the  general transition GPDs when the related complete experimental data are available.  The CLAS data on this transition is ongoing \cite{Kim:2022bwn,Proceedings:2020fyd}.

\begin{figure}[htp]
\centering
 \subfloat[]{\includegraphics[width=0.33\textwidth]{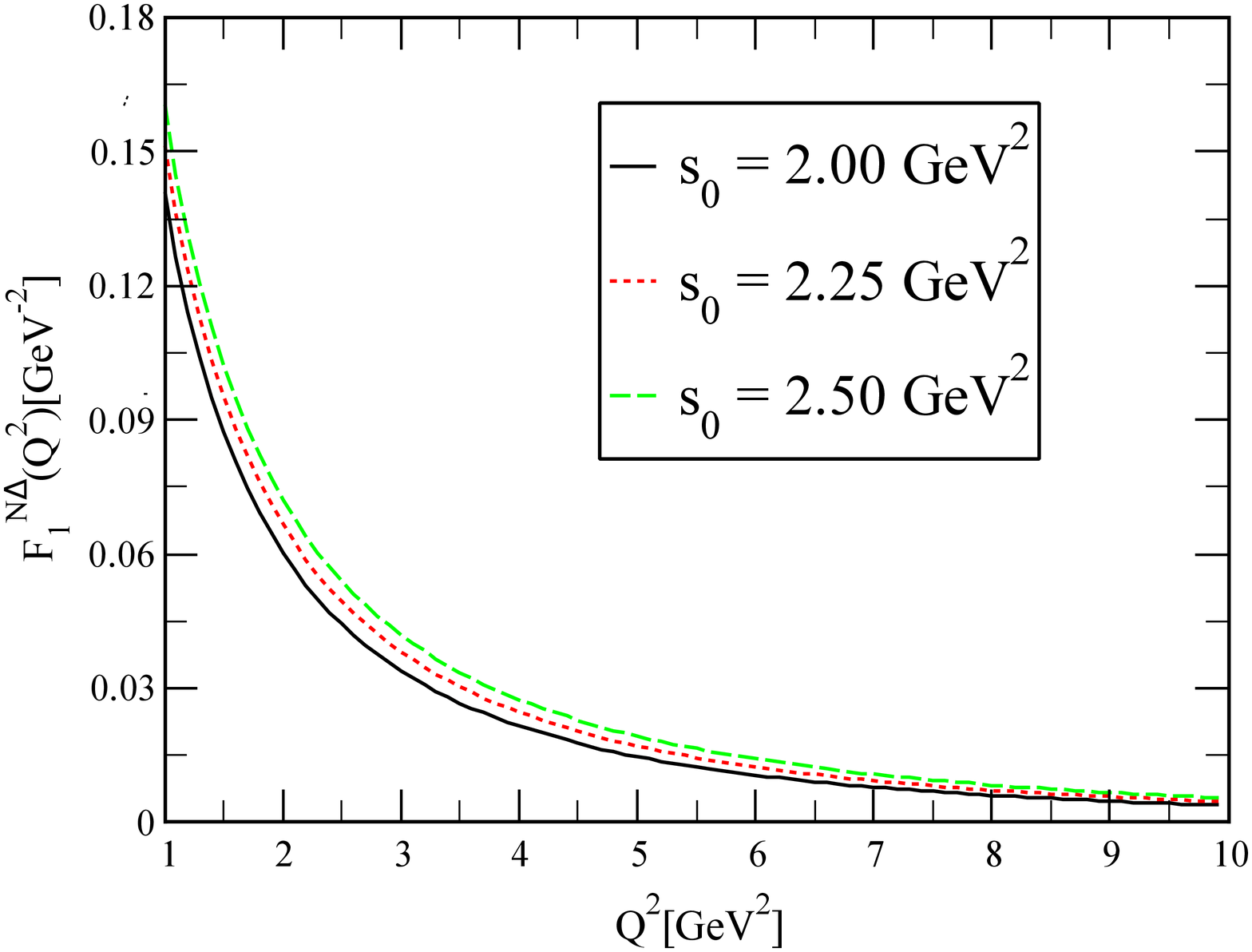}}~~~~~~~~~~
 \subfloat[]{\includegraphics[width=0.33\textwidth]{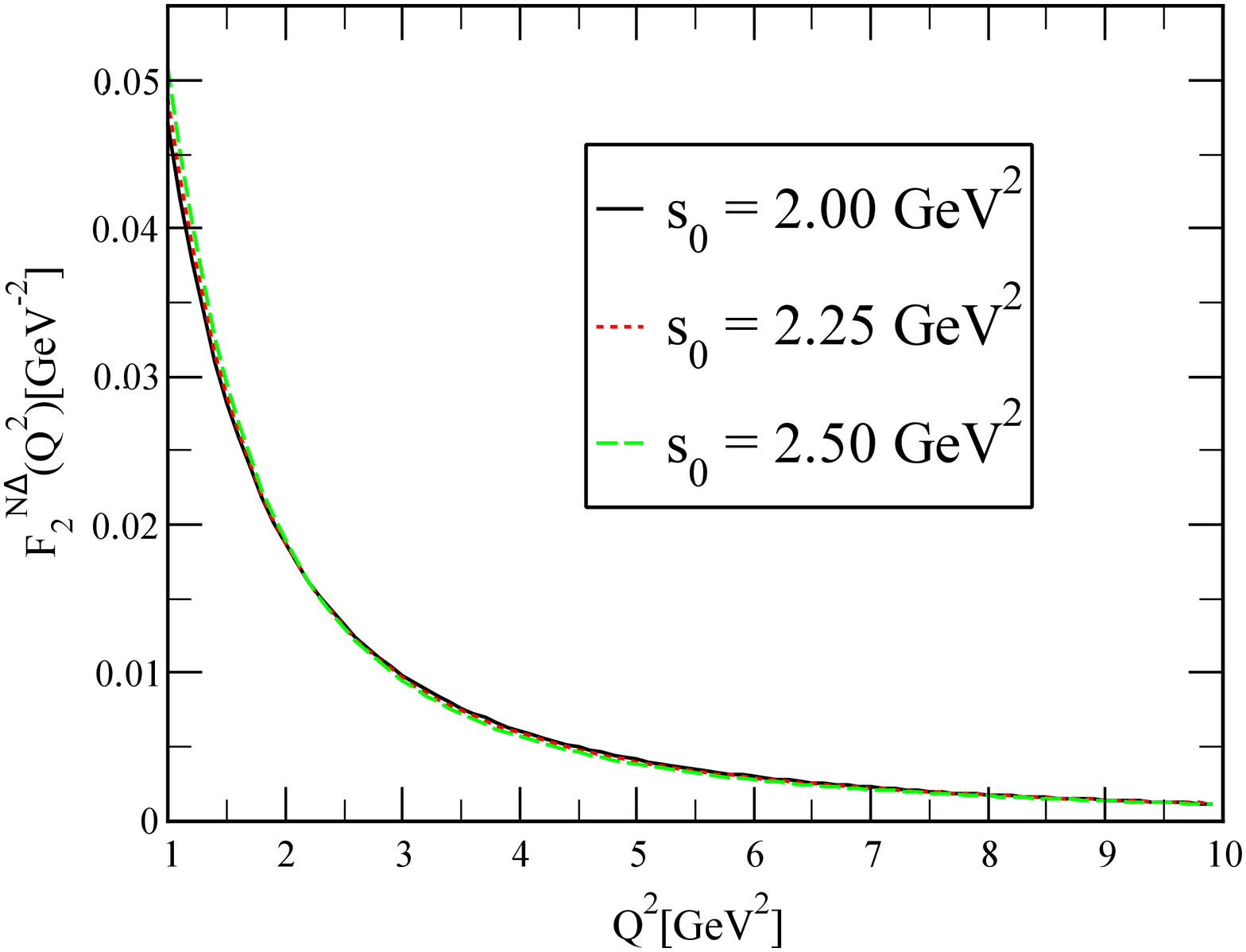}}\\
 \subfloat[]{\includegraphics[width=0.33\textwidth]{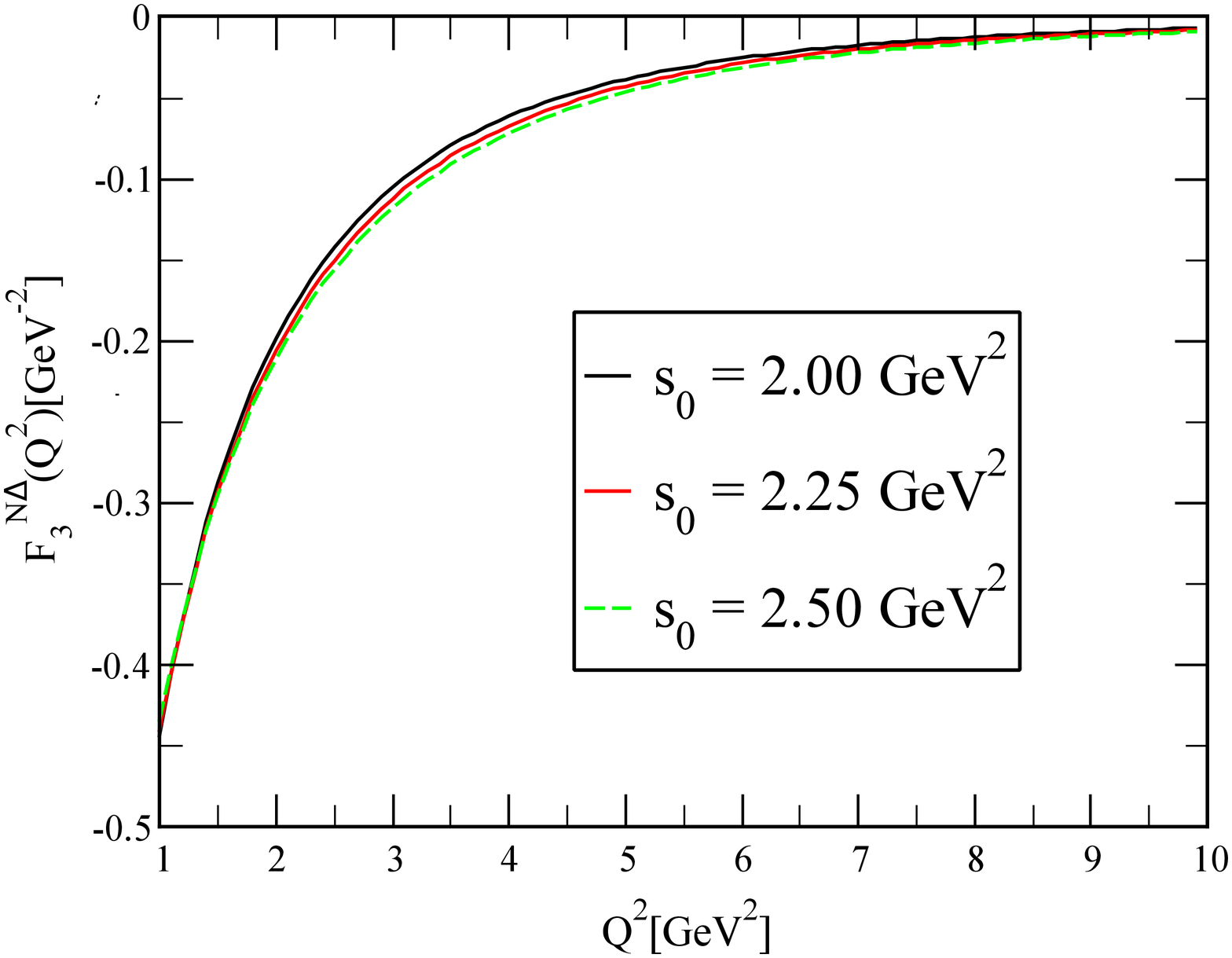}}~~~~~~~~~~
 \subfloat[]{\includegraphics[width=0.33\textwidth]{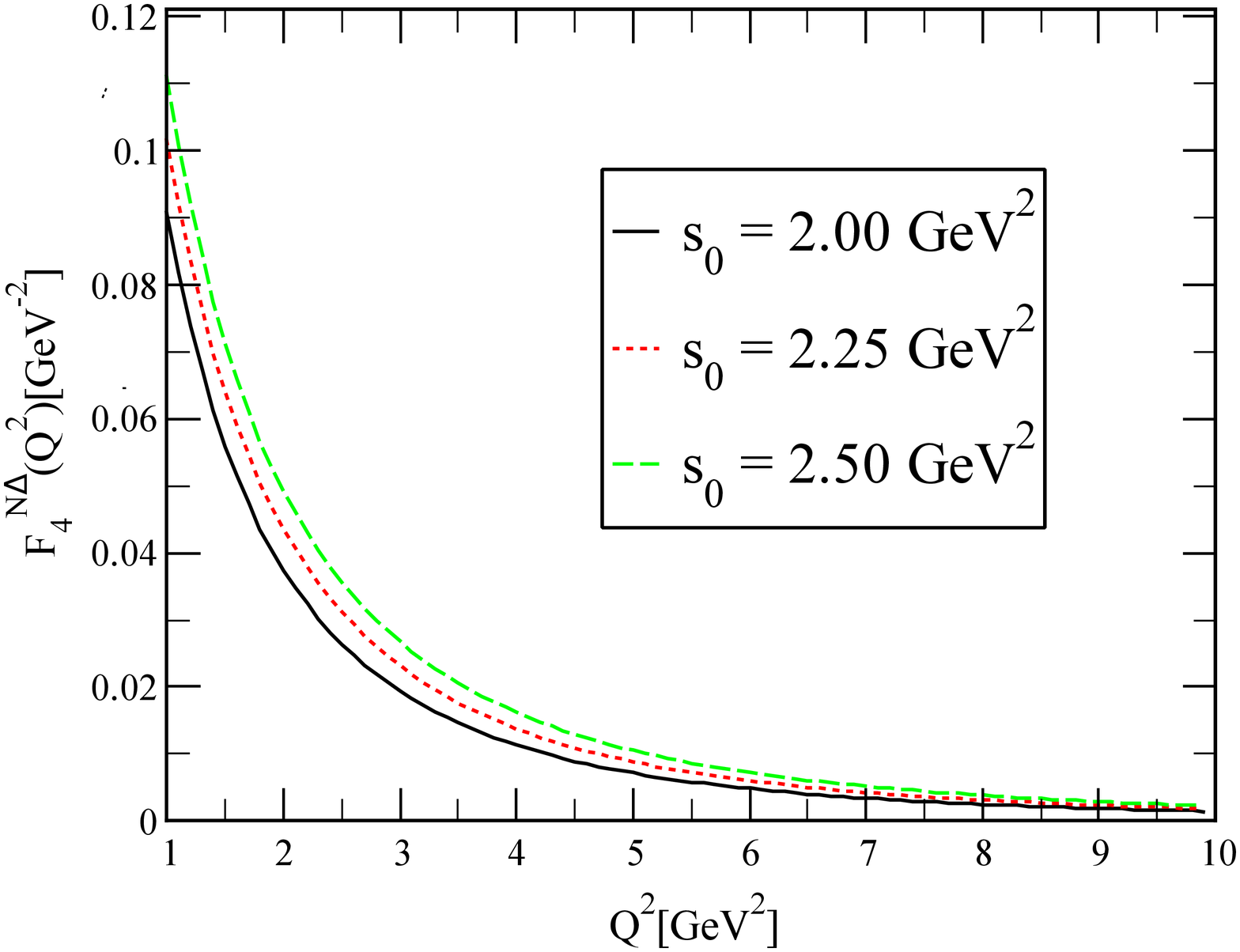}}\\
 \subfloat[]{\includegraphics[width=0.33\textwidth]{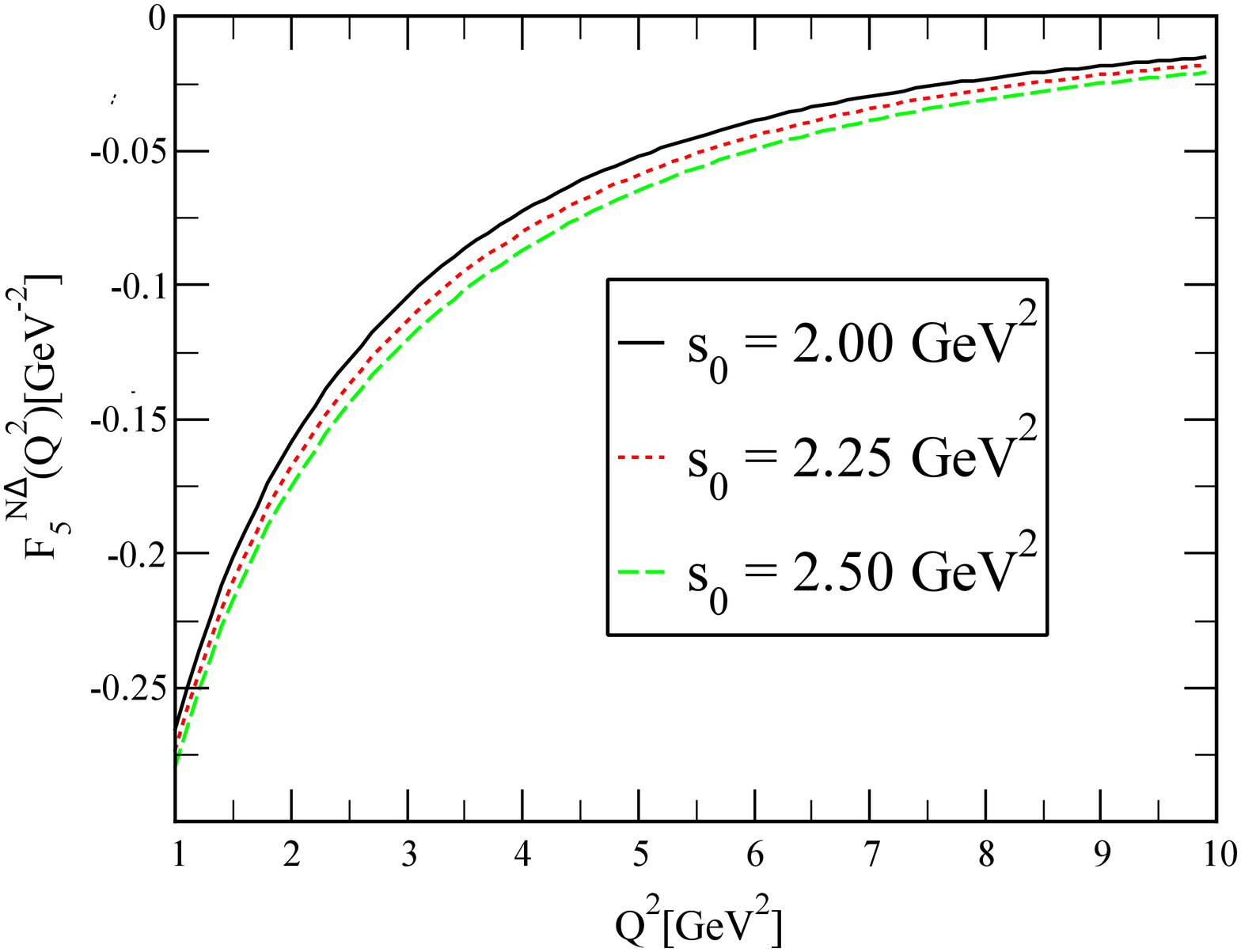}}~~~~~~~~~~
 \subfloat[]{\includegraphics[width=0.33\textwidth]{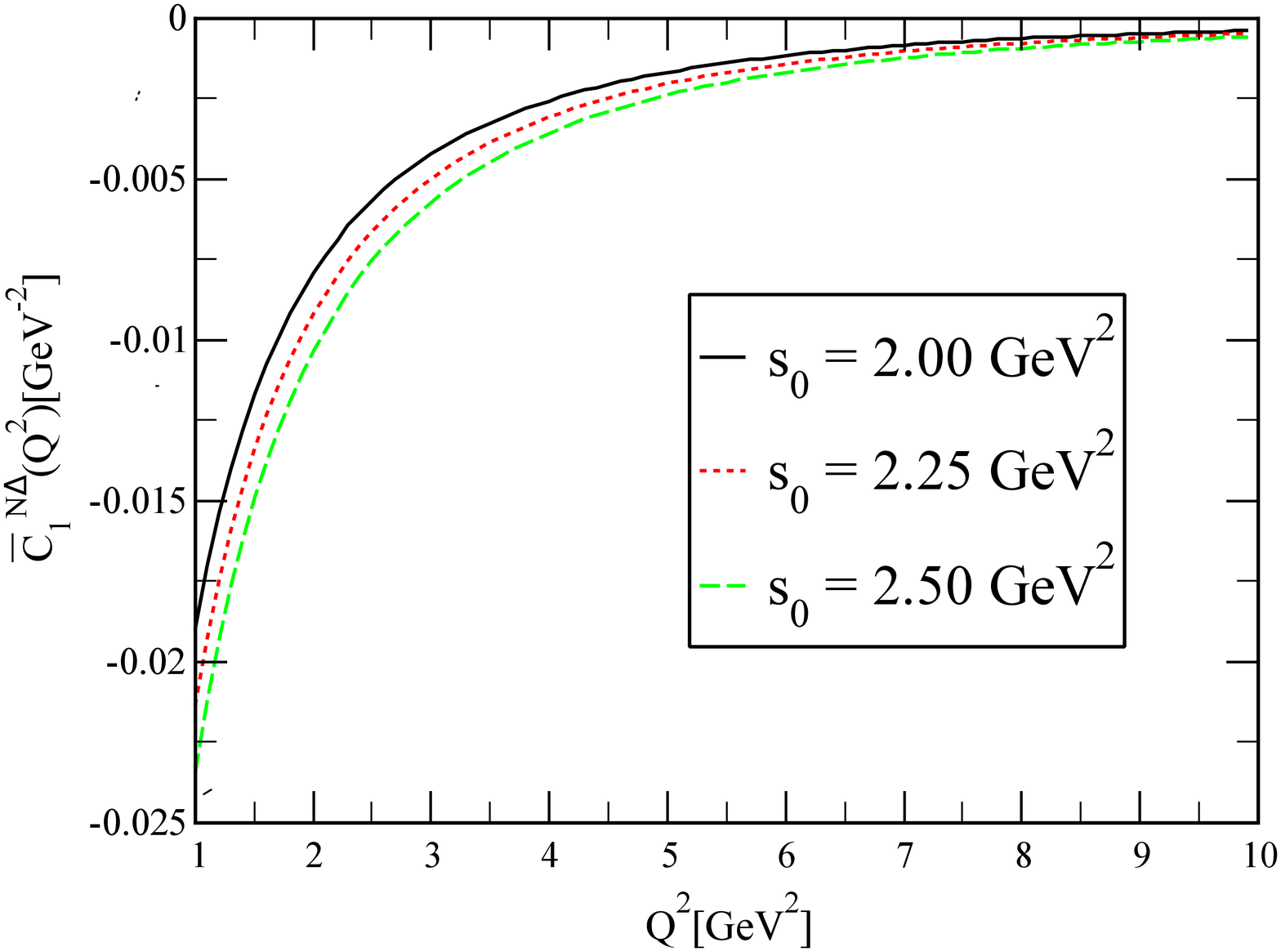}}\\
 \subfloat[]{\includegraphics[width=0.33\textwidth]{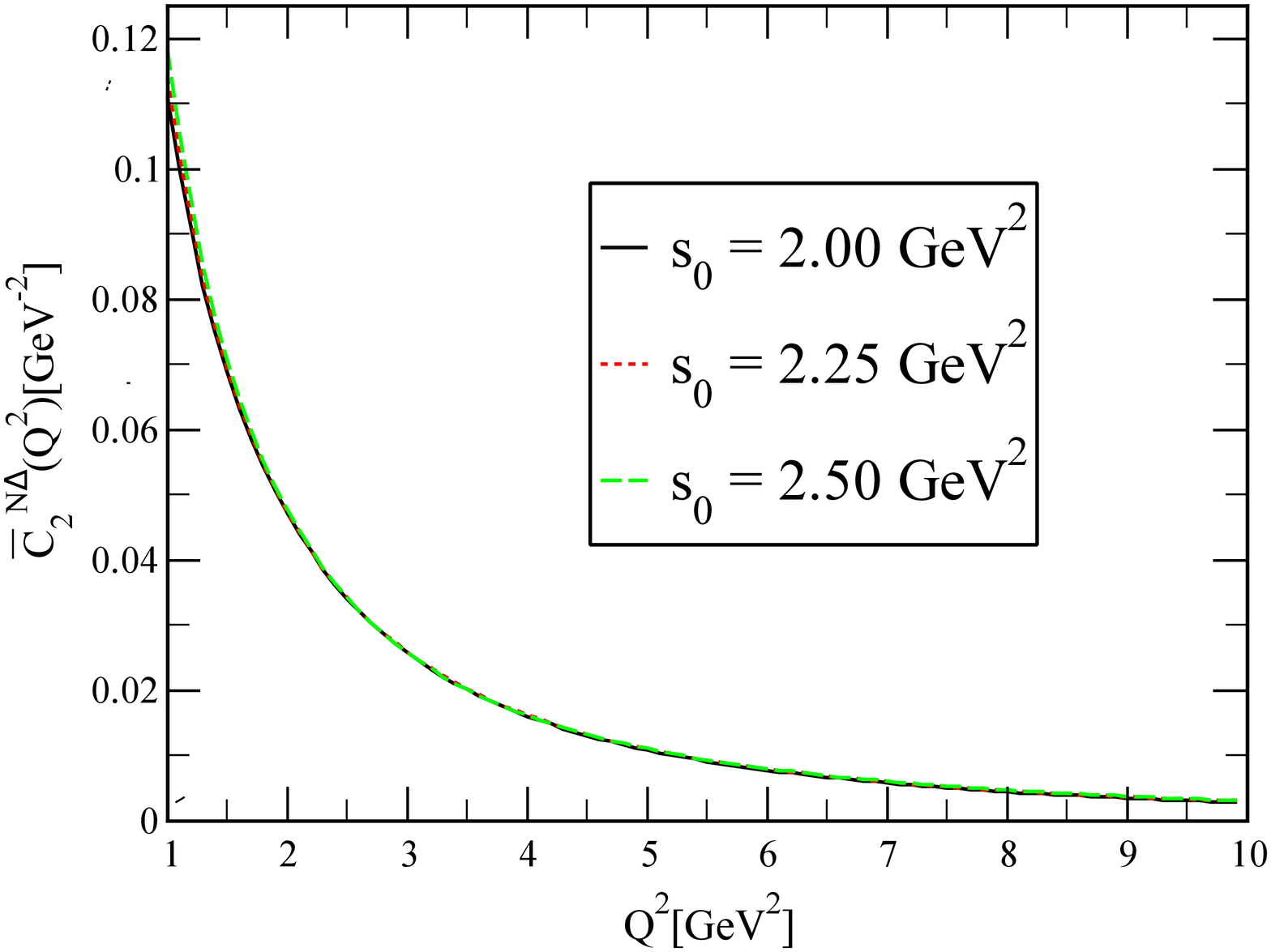}}~~~~~~~~~~
 \subfloat[]{\includegraphics[width=0.33\textwidth]{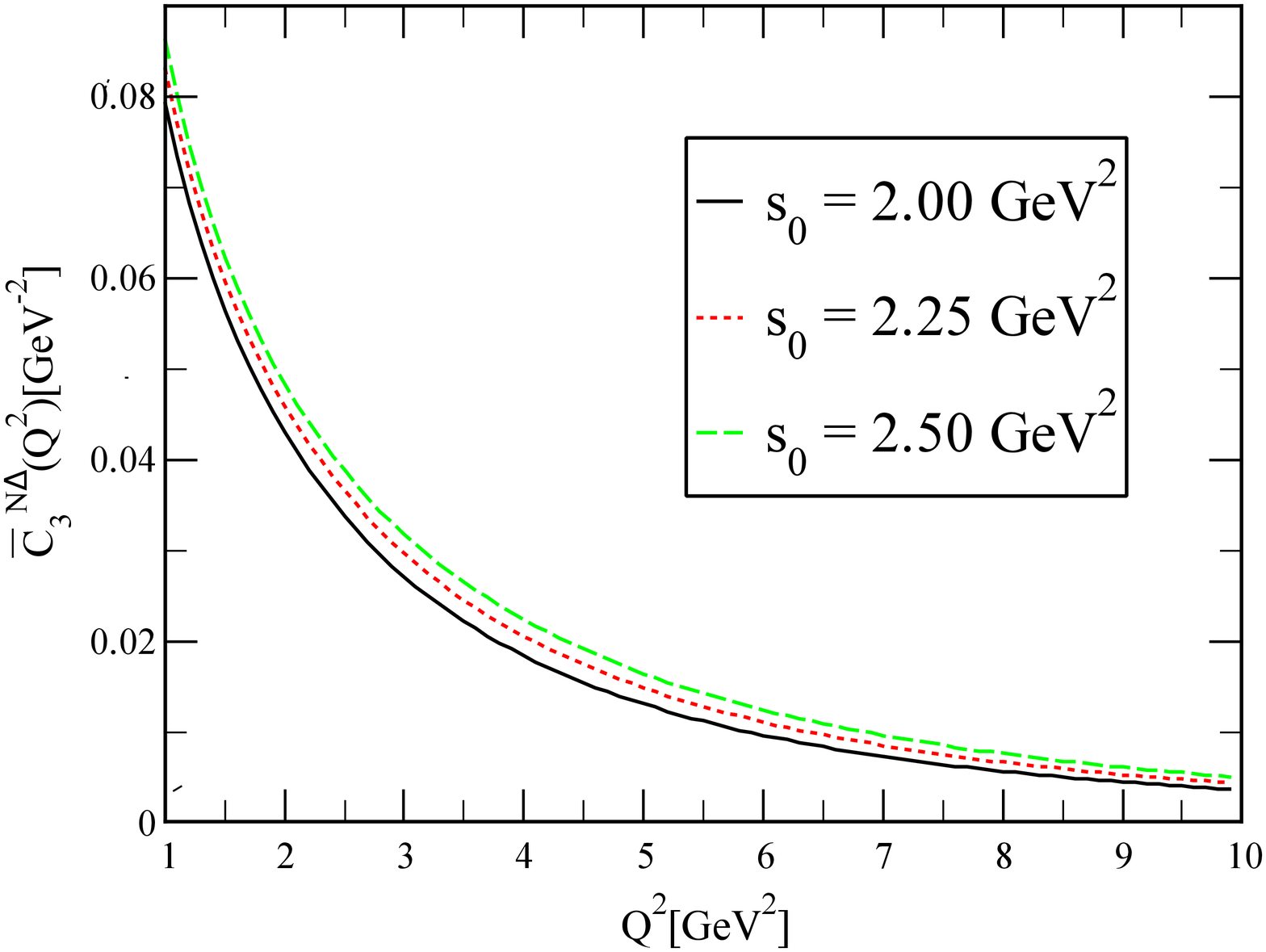}}
 \caption{The dependence of the $N \rightarrow \Delta$ transition GFFs on $Q^2$ at fixed values of the  $s_0$,  average $M^2$  and   set-I parameters.}
 \label{Qsqfigs1}
  \end{figure}

\begin{figure}[htp]
\centering
 \subfloat[]{\includegraphics[width=0.33\textwidth]{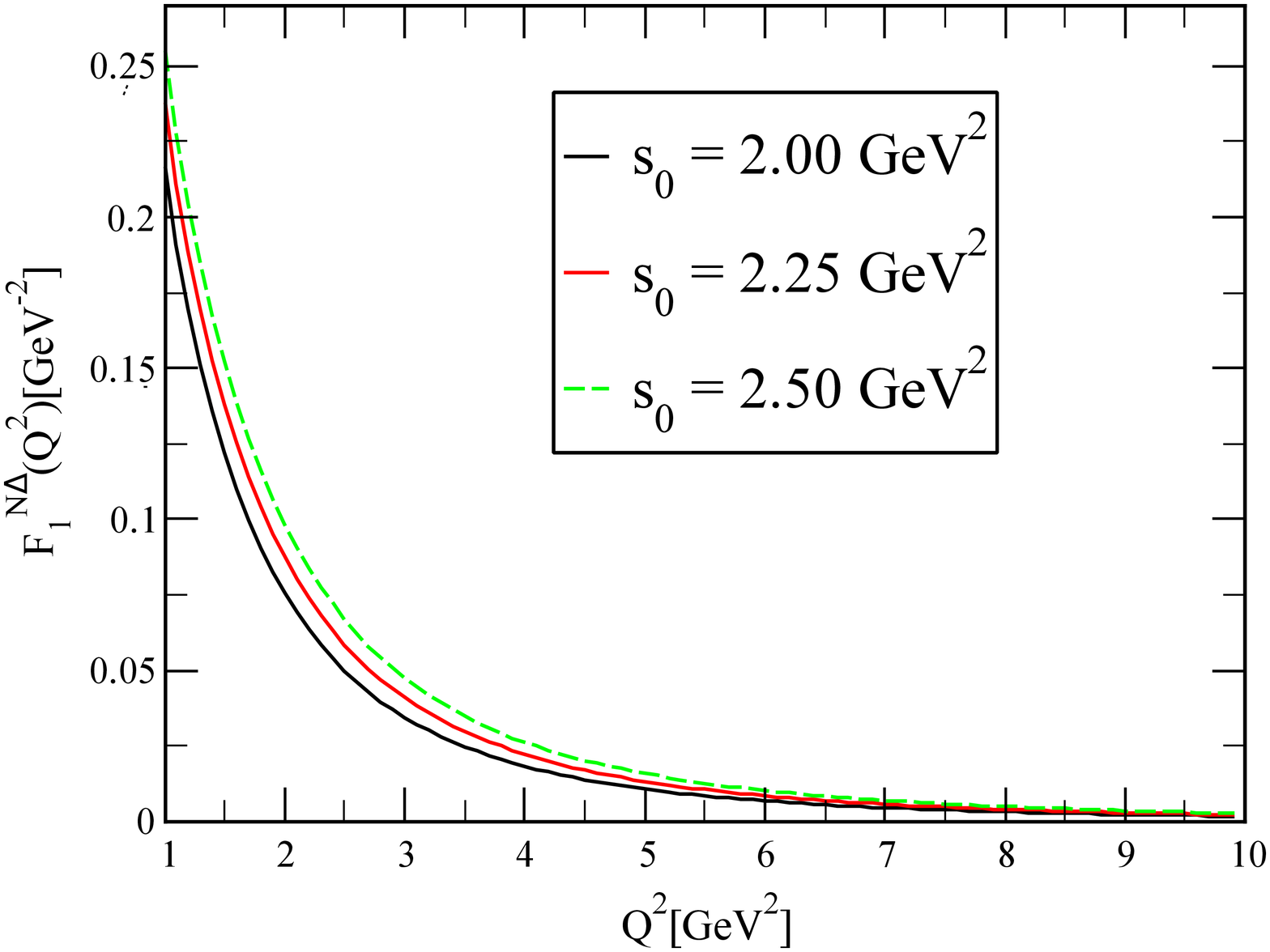}}~~~~~~~~~~
 \subfloat[]{\includegraphics[width=0.33\textwidth]{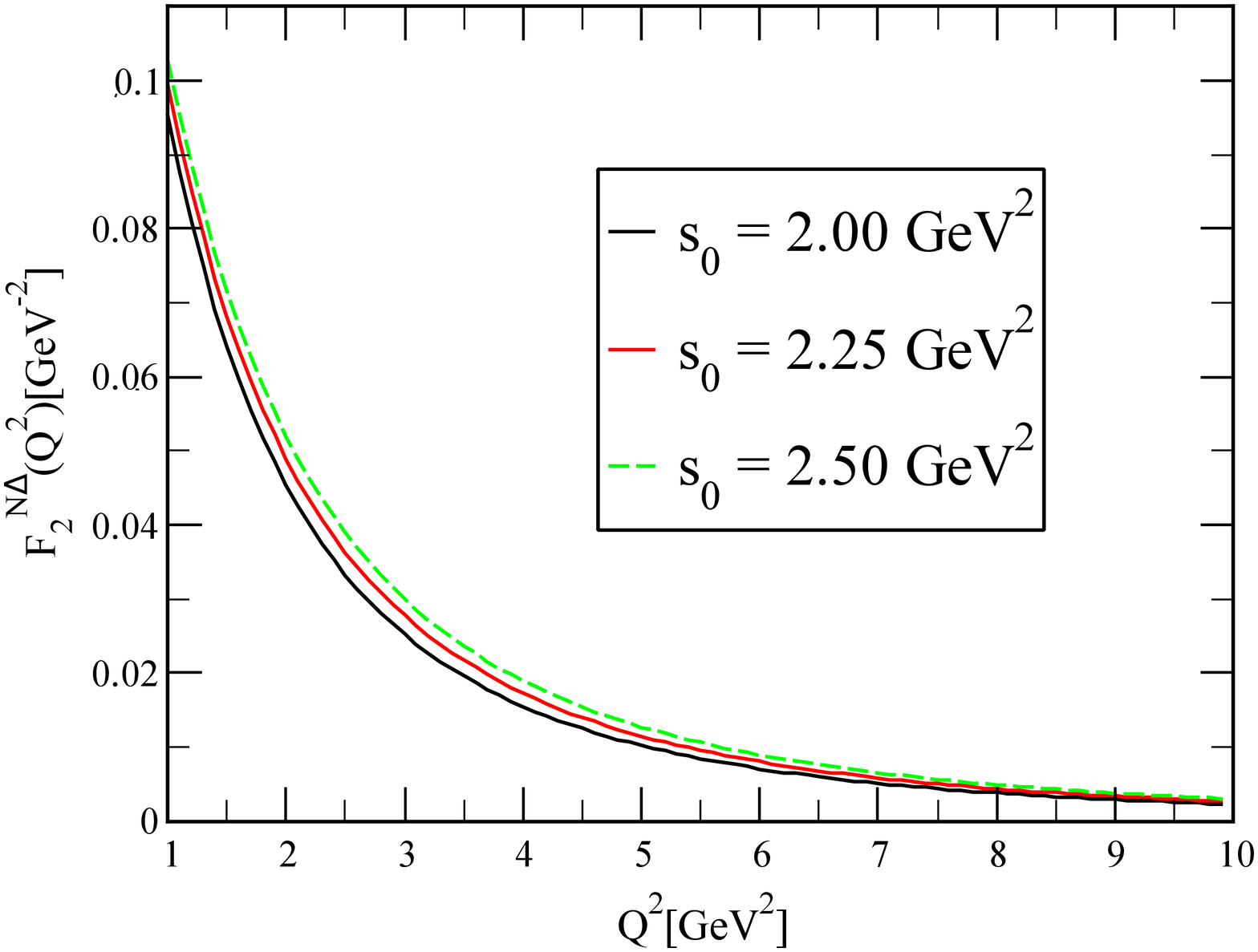}}\\
 \subfloat[]{\includegraphics[width=0.33\textwidth]{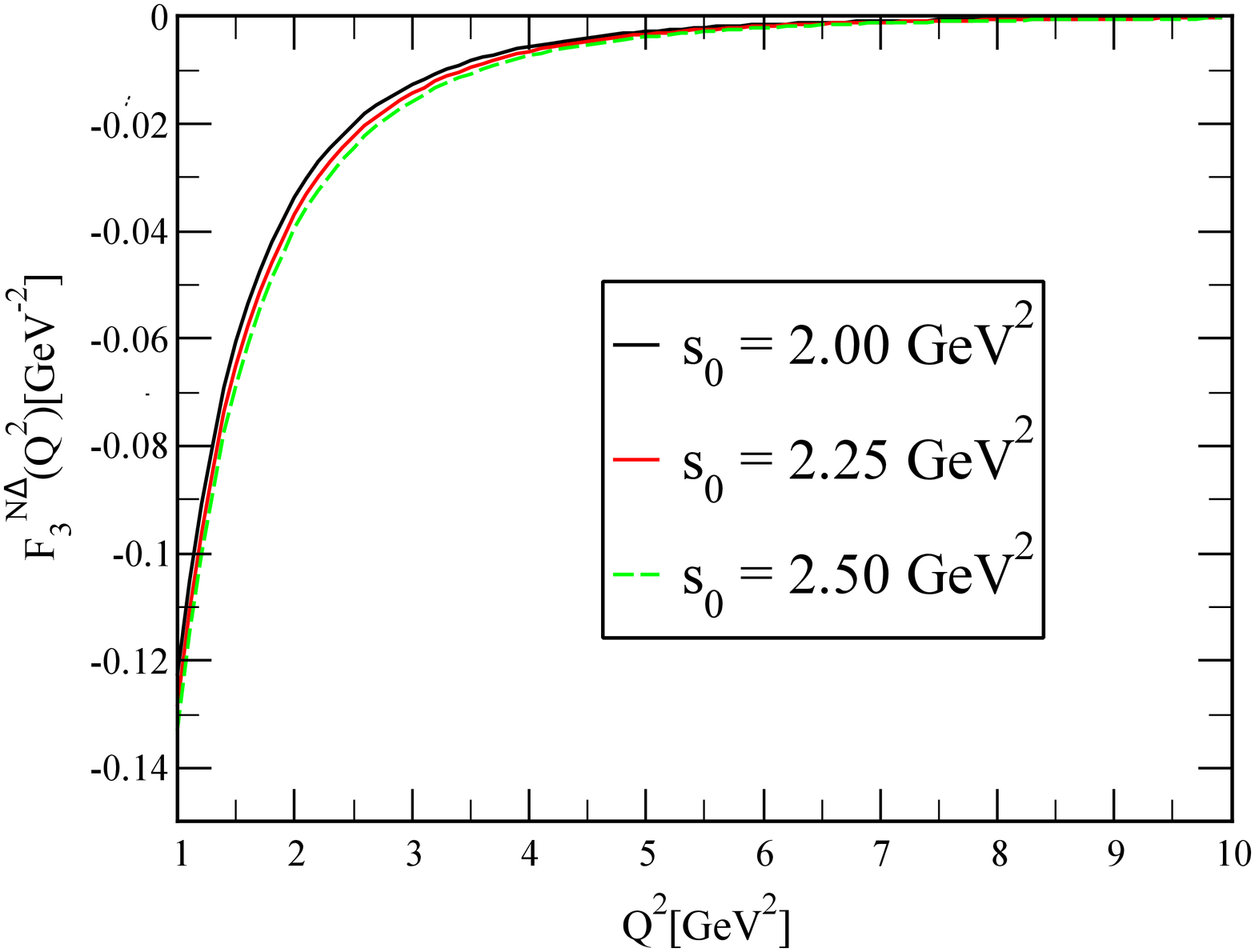}}~~~~~~~~~~
 \subfloat[]{\includegraphics[width=0.33\textwidth]{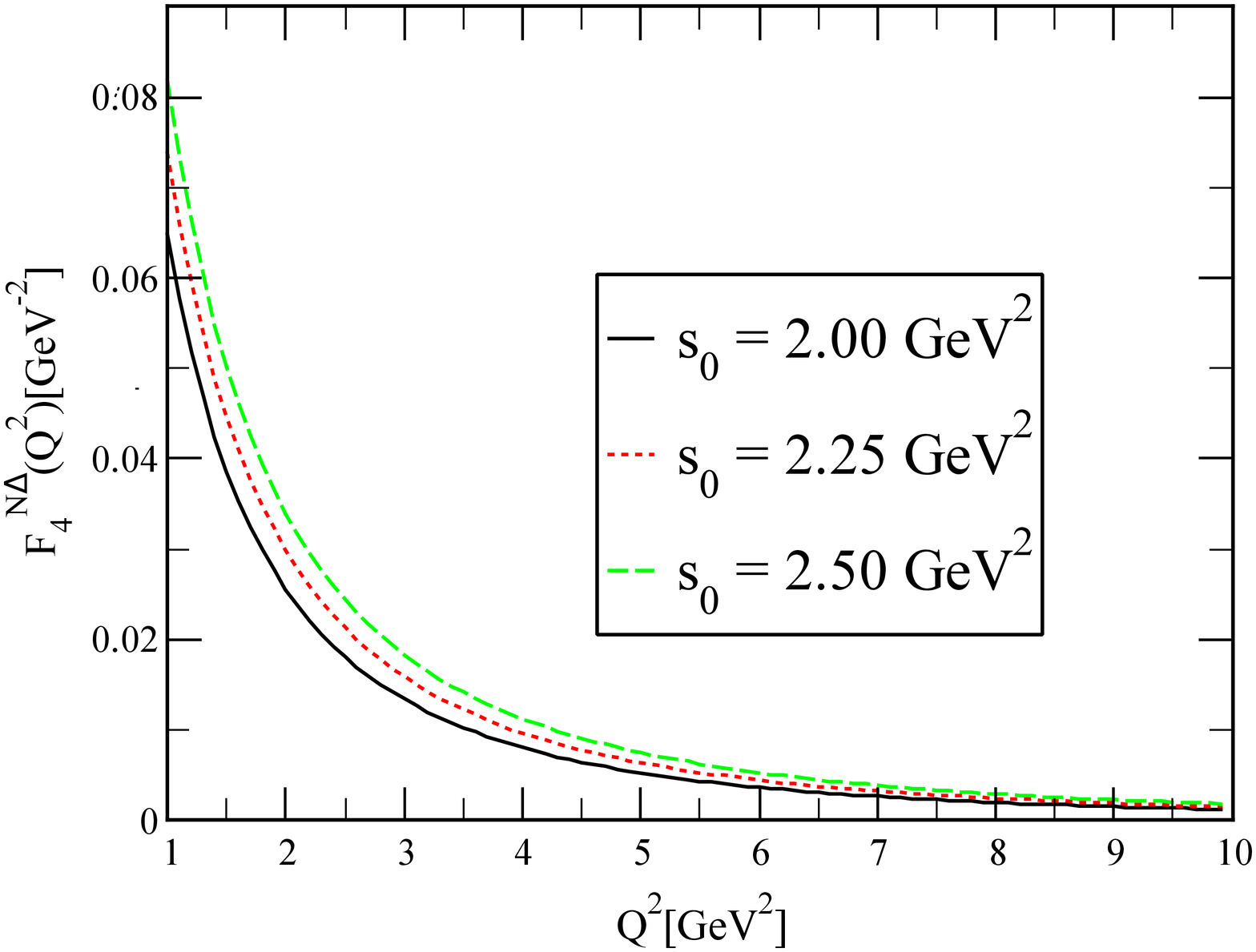}}\\
 \subfloat[]{\includegraphics[width=0.33\textwidth]{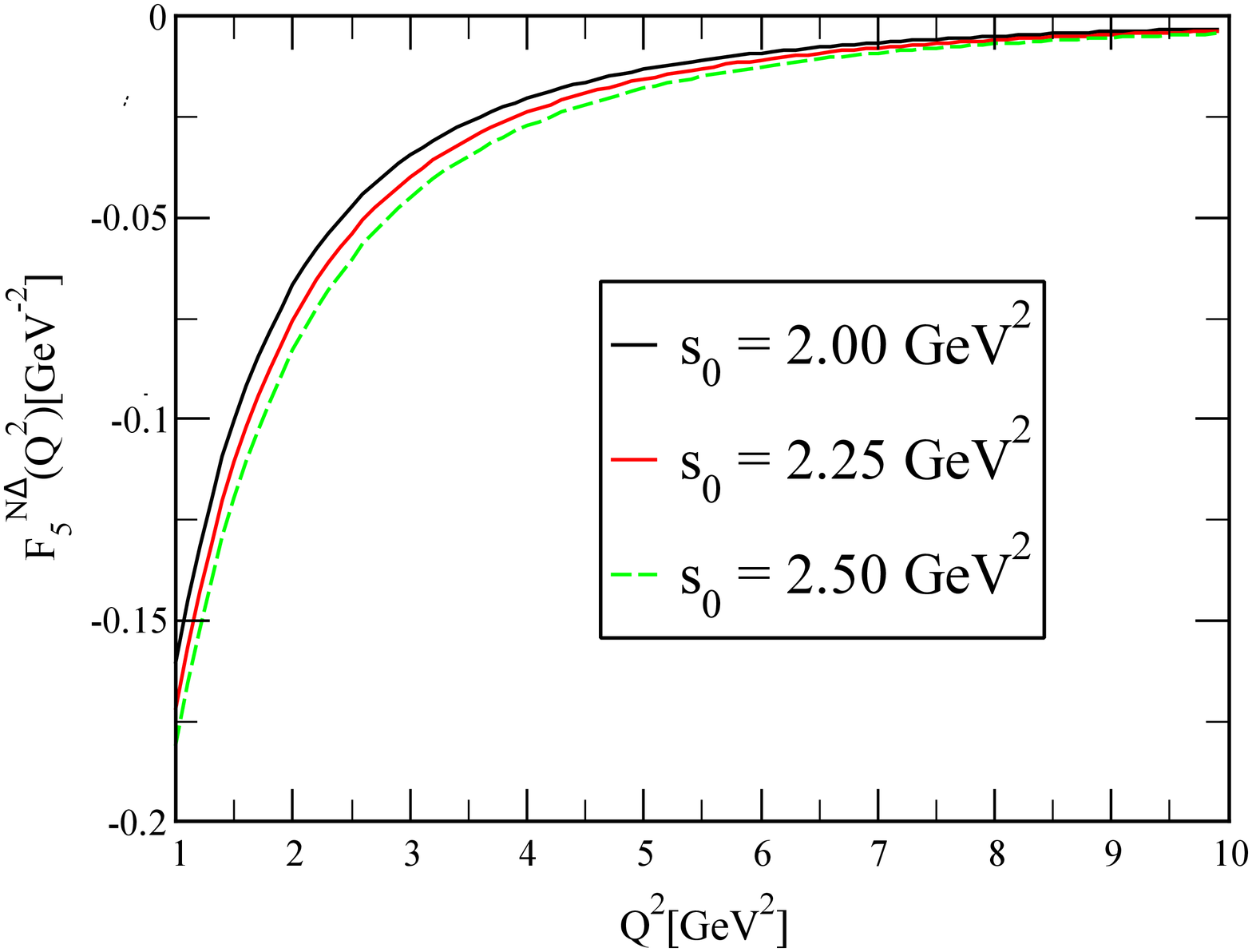}}~~~~~~~~~~
 \subfloat[]{\includegraphics[width=0.33\textwidth]{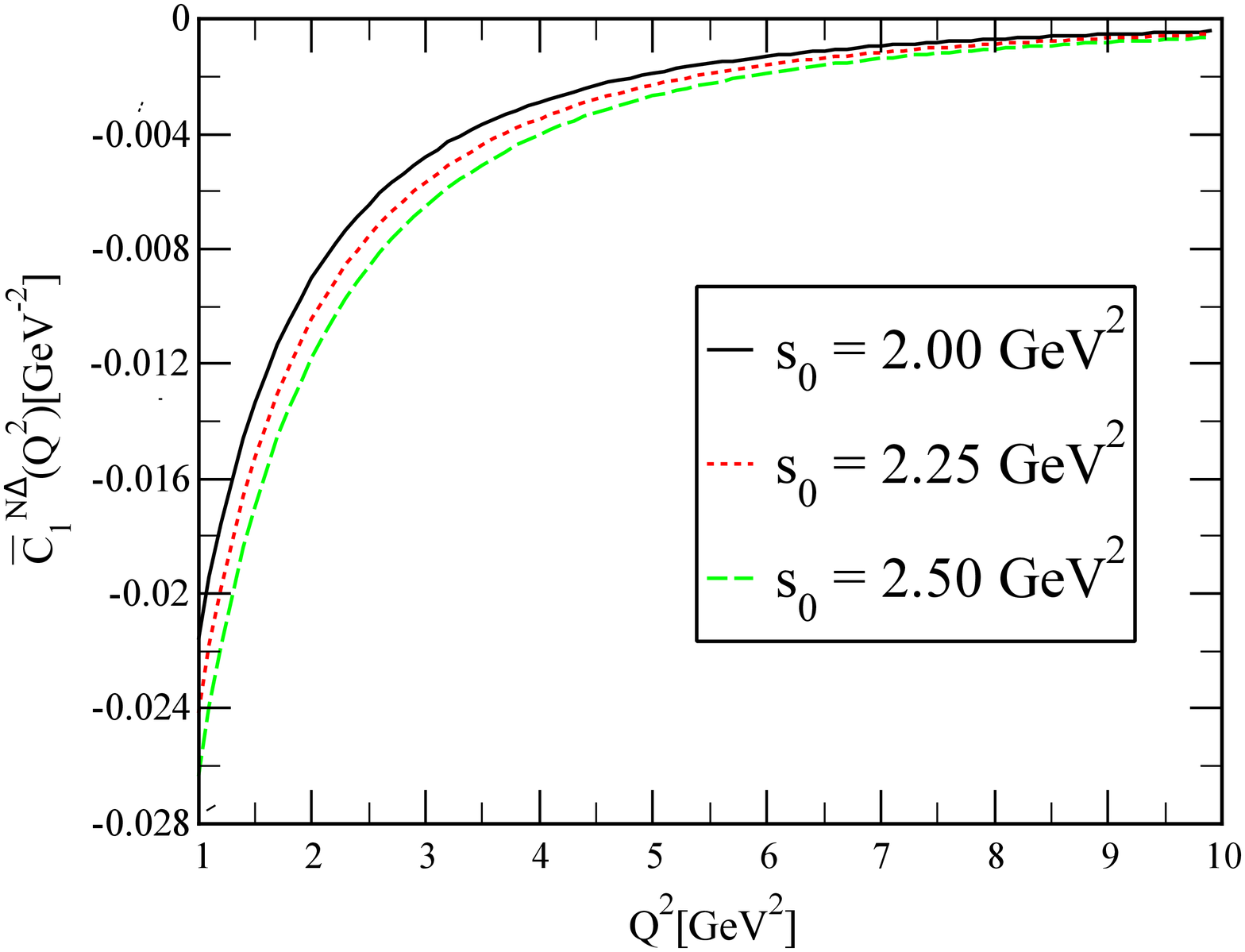}}\\
 \subfloat[]{\includegraphics[width=0.33\textwidth]{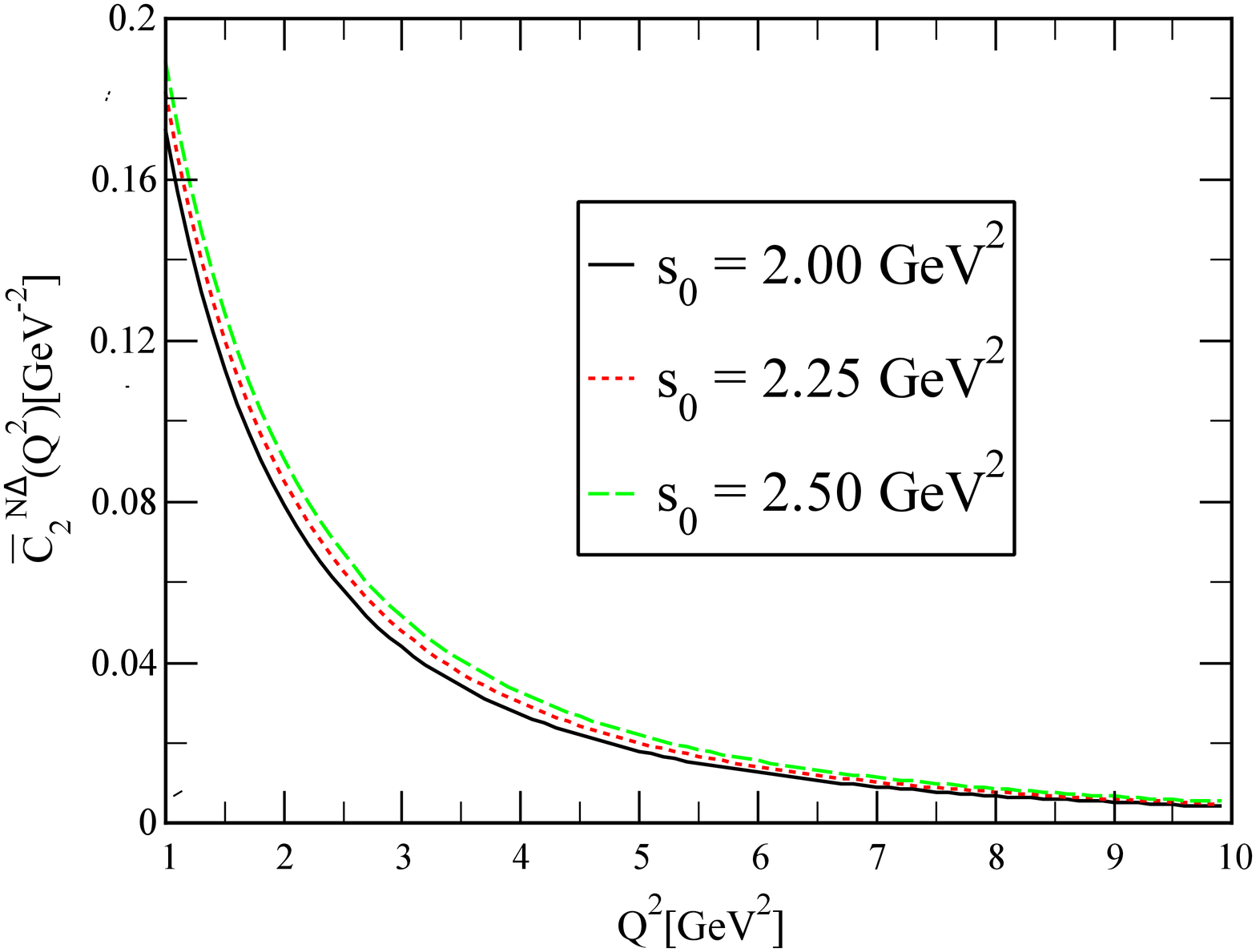}}~~~~~~~~~~
 \subfloat[]{\includegraphics[width=0.33\textwidth]{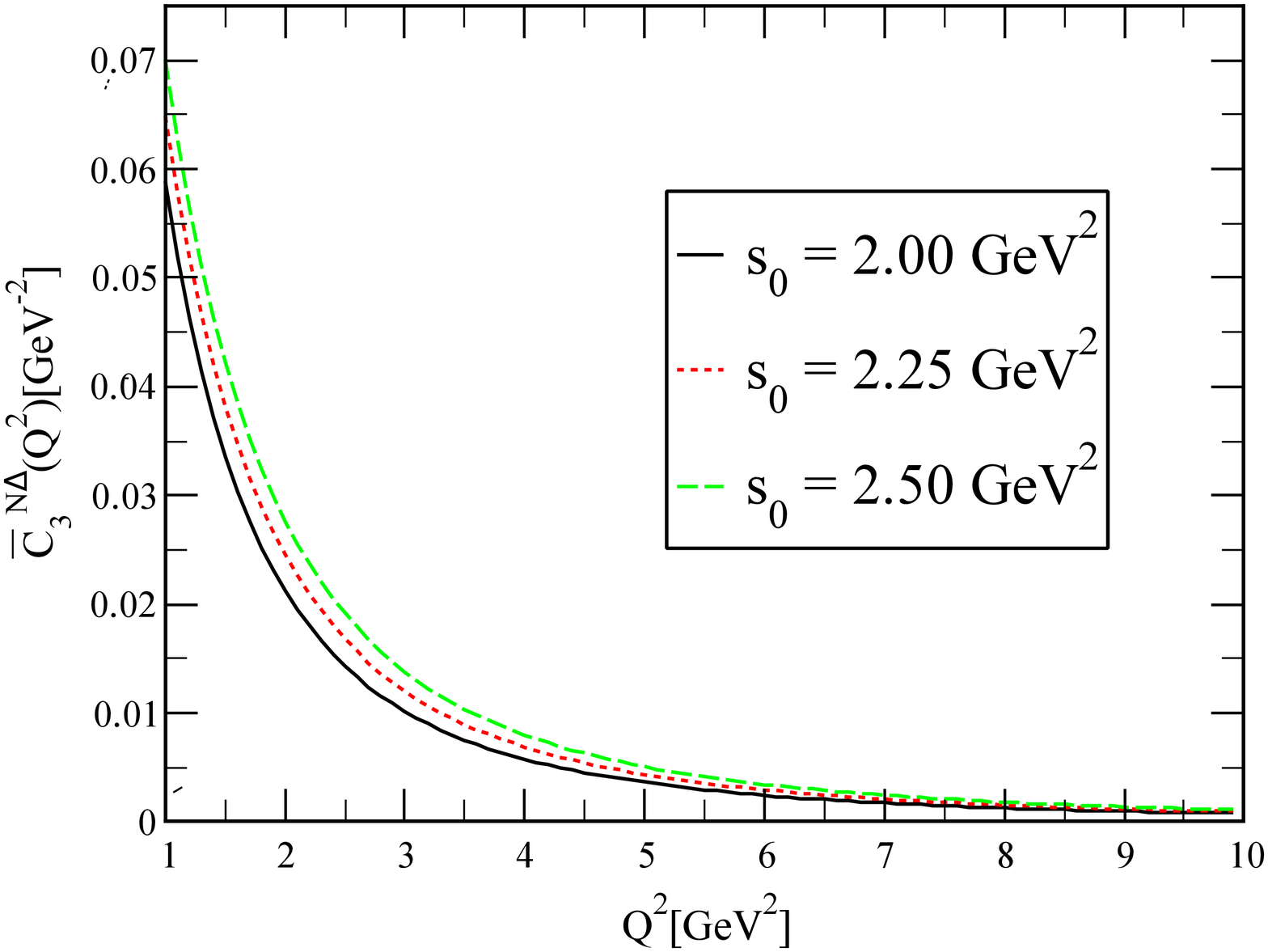}}
 \caption{The dependence of the $N \rightarrow \Delta$ transition GFFs on $Q^2$ at fixed values of the  $s_0$, average  $M^2$  and set-II parameters.}
 \label{QsqfigsII}
  \end{figure}

 Using the above fit functions, the $Q^2 $-behavior of the GFFs in a wide range, $ Q^2\in[0-10]$ ~GeV$^2$, are depicted in figures \ref{Qsqfigs1} and  \ref{QsqfigsII}. Our results may be checked by future related experiments, Lattice QCD and other phenomenological models.


\section{Summary and Concluding Remarks}\label{secIV}

We investigated the transition GFFs of the $N \rightarrow \Delta$. The quark part of the EMT current are parameterized in terms of 
nine independent  (five conserved and four  non-conserved) form factors. To calculate these GFFs, we applied the light-cone QCD sum rule technique using the DAs of the on-shell nucleon. The nine independent Lorentz structures entering to the calculations in both the hadronic and QCD sides allowed us to extract the sum rules for the desired GFFs. We numerically analyzed the obtained sum rules for two sets of input parameters inside the DAs of the nucleon. We used multipole fit functions to extrapolate the results to the small values of $Q^2$, $0\leq Q^2<1$~GeV$^2$, to find the values of the GFFs at static limit. We presented the $Q^2$-behaviors of the form factors in the interval  $ [0-10]$~GeV$^2$, using the working windows of the auxiliary parameters. 

As we previously mentioned, investigation of the EMT current interactions of hadrons besides their electromagnetic, weak and strong interactions can give us valuable information about their mass and  spin as well as the pressure and shear force their inside. The transition GFFs of $N \rightarrow \Delta$ calculated in the present study include useful knowledge on the $N-\Delta$ system. Our results can be checked in future experiments. Comparison of our results with the future Lattice QCD and other phenomenological predictions will be of great importance as well.
The direct measurement of the $N-\Delta$  GFFs may not be possible with the present facilities. However, in principle, these GFFs can be extracted from the $N-\Delta$ GPDs by considering them  as the  second Mellin moment of the transition GPDs. The  project of extracting the transition GPDs from the CLAS data is ongoing \cite{Kim:2022bwn,Proceedings:2020fyd}.
  
\section*{ACKNOWLEDGEMENTS}
K. Azizi is grateful to Iran Science Elites Federation (Saramadan)
for the partial  financial support provided under the grant number ISEF/M/401385.

\appendix
\section*{Appendix: Explicit forms of the $\rho_i^{QCD}(v,y,x_1,x_2,x_3)$ functions}

In this appendix, we present the explicit expressions for the $\rho_i^{QCD}(v,y,x_1,x_2,x_3)$ functions:
\begin{align}
 \rho_1^{QCD}(v,y,x_1,x_2,x_3)&=
 \frac{m_N^3}{2\sqrt{3}}\int_0^1 \frac{(1+x_2)}{(q-p x_2)^4}dx_2 \int_0^{1-x_2}d{x_1} \big[\big(    A_1^M-T_1^M-2V_1^M\big)(x_1,x_2,1-x_1-x_2)\big]\nonumber\\
 &-\frac{m_N^3}{2\sqrt{3}}\int_0^1 \frac{(1+x_3)}{(q-p x_3)^4}dx_3 \int_0^{1-x_3}d{x_1} \big[T_1^M(x_1,1-x_1-x_3,x_3)\big]\nonumber\\
 &-\frac{m_N}{4 \sqrt{3}}\int_0^1 \frac{(1+x_2)}{(q-p x_2)^2}dx_2 \int_0^{1-x_2}d{x_1} \big[\big( 2A_1+A_3-2V_1+V_3\big)(x_1,x_2,1-x_1-x_2)\big]\nonumber\\
 &+\frac{m_N}{4 \sqrt{3}}\int_0^1 \frac{(1+x_3)}{(q-p x_3)^2}dx_3 \int_0^{1-x_3}d{x_1} \big[\big( A_3+2T_1+V_3\big)(x_1,1-x_1-x_3,x_3)\big]\nonumber\\
 &+\frac{m_N^3}{4 \sqrt{3}}\int_0^1 \frac{(1+v)}{(q-p v)^4} dv \int_0^v dy \int_y^1 d{x_2} \int_0^{1-x_2} d{x_1} \big[\big(T_1-T_2-T_5+T_6 -2T_7-2T_8\nonumber\\
 &+V_1-V_2-V_3-V_4-V_5+V_6\big)(x_1,x_2,1-x_1-x_2)\big]
 \nonumber\\
 &+\frac{m_N^3}{4 \sqrt{3}}\int_0^1 \frac{(1+v)}{(q-p v)^4} dv \int_0^v dy \int_y^1 d{x_3} \int_0^{1-x_3} d{x_1}\big[\big(-A_1+A_2-A_3-A_4+A_5-A_6-4T_1 \nonumber\\
 &+2T_2+2T_3+2T_4+2T_5-4T_6+2T_7+2T_8+V_1-V_2-V_3-V_5+V_6\big)(x_1,1-x_1-x_3,x_3)\big]
 \nonumber\\
 & +\frac{m_N}{4\sqrt{3}}\int_0^1 \frac{1}{(q-p y)^2} dy \int_y^1 d{x_2} \int_0^{1-x_2}d{x_1} \big[\big(-2A_1+2A_2-2A_3-T_1+T_2+2T_7+2V_1-2V_2  \nonumber\\& 
 -2V_3  \big)(x_1,x_2,1-x_1-x_2)\big]\nonumber\\
 &+\frac{m_N}{\sqrt{3}}\int_0^1 \frac{1}{(q-p y)^2} dy \int_y^1 d{x_3} \int_0^{1-x_3}d{x_1} \big[\big(T_1+T_3-T_7 \big)(x_1,1-x_1-x_3,x_3)\big],\\
 \nonumber\\
 \rho_2^{QCD}(v,y,x_1,x_2,x_3)&= \frac{1}{\sqrt{3}}\int_0^1 \frac{x_2}{(q-p x_2)^2}dx_2 \int_0^{1-x_2}d{x_1} \big[\big( -A_1+V_1\big)(x_1,x_2,1-x_1-x_2)\big]\nonumber\\
 &+\frac{2}{ \sqrt{3}}\int_0^1 \frac{x_3}{(q-p x_3)^2}dx_3 \int_0^{1-x_3}d{x_1} \big[T_1(x_1,1-x_1-x_3,x_3)\big]
 \nonumber\\
 &+ \frac{m_N^2}{\sqrt{3}}\int_0^1 \frac{x_2^2}{(q-p x_2)^4}dx_2 \int_0^{1-x_2}d{x_1} \big[\big(    A_1^M+4T_1^M-2V_1^M\big)(x_1,x_2,1-x_1-x_2)\big]\nonumber\\
 &-\frac{m_N^2}{\sqrt{3}}\int_0^1 \frac{x_3^2}{(q-p x_3)^4}dx_3 \int_0^{1-x_3}d{x_1} \big[\big( -A_1^M-2T_1^M+V_1^M\big)(x_1,1-x_1-x_3,x_3)\big]\nonumber\\
  &
 +\frac{ m_N^2}{\sqrt{3}}\int_0^1 \frac{y^2}{(q-p y)^4} dy \int_y^1 d{x_2} \int_0^{1-x_2}d{x_1}\big[\big(-A_1+A_2-A_3+V_1-V_2-V_3\big)\nonumber\\
 &\times (x_1,x_2,1-x_1-x_2)\big]\nonumber\\
 &
 +\frac{ m_N^2}{2\sqrt{3}}\int_0^1 \frac{y^2}{(q-p y)^4} dy \int_y^1 d{x_3} \int_0^{1-x_3}d{x_1}\big[\big(A_3-A_4-4T_1-2T_3-2T_5-4T_7\nonumber\\
 &-4T_8-V_1+V_4+V_5\big)(x_1,1-x_1-x_3,x_3)\big]\nonumber
 \end{align}
\begin{align}
&+\frac{2m_N^2}{\sqrt{3}}\int_0^1 \frac{v^2}{(q-p v)^4} dv \int_0^v dy \int_y^1 d{x_2} \int_0^{1-x_2} d{x_1} \big[\big(-T_2+T_3+T_5-T_5-T_7-T_8\big)\nonumber\\
 &\times(x_1,x_2,1-x_1-x_2)\big]
 \nonumber\\
 &+\frac{2m_N^2}{\sqrt{3}}\int_0^1 \frac{v^2}{(q-p v)^4} dv \int_0^v dy \int_y^1 d{x_3} \int_0^{1-x_3} d{x_1}\big[\big(-T_2+T_3+T_5-T_5-T_7-T_8\big) \nonumber\\
 &\times (x_1,1-x_1-x_3,x_3)\big],\\
 \nonumber\\
 \rho_3^{QCD}(v,y,x_1,x_2,x_3)&=\frac{ m_N}{\sqrt{3}}\int_0^1 \frac{(1+2y+y^2)}{(q-p y)^4} dy \int_y^1 d{x_3} \int_0^{1-x_3}d{x_1}\big[\big(A_1-A_2+A_3-V_1+V_2+V_3+2T_1-2T_3\nonumber\\
 &-2T_7\big)  (x_1,1-x_1-x_3,x_3)\big]\nonumber\\
 &
 +\frac{2m_N^3}{ \sqrt{3}}\int_0^1 \frac{v(1+v+v^2)}{(q-p v)^6} dv \int_0^v dy \int_y^1 d{x_3} \int_0^{1-x_3} d{x_1}\big[\big(A_1-A_2+A_3+A_4-A_5+A_6+2T_1 \nonumber\\
 &-2T_3-2T_4+2T_6-2T_7-2T_8-V_1+V_2+V_3+V_5-V_6\big)(x_1,1-x_1-x_3,x_3)\big],\\
 \nonumber\\
%
%
 \rho_4^{QCD}(v,y,x_1,x_2,x_3)&=\frac{ m_N^2}{2\sqrt{3}}\int_0^1 \frac{1}{(q-p y)^2} dy \int_y^1 d{x_2} \int_0^{1-x_2}d{x_1}\big[\big(A_1-A_2+A_4+P_1-P_2-S_1+S_2+T_1-T_2\nonumber\\
 &-T_7-V_1+V_2+V_4\big) (x_1,x_2,1-x_1-x_2)\big]\nonumber\\
 &
 +\frac{ m_N^2}{4\sqrt{3}}\int_0^1 \frac{1}{(q-p y)^2} dy \int_y^1 d{x_3} \int_0^{1-x_3} d{x_1}
 \big[\big(A_3-A_4-4T_1+4T_3+4T_7+V_3-V_4\big) \nonumber\\
 &\times(x_1,1-x_1-x_3,x_3)\big],
 \\
 \nonumber\\
 \rho_5^{QCD}(v,y,x_1,x_2,x_3)&=\frac{ 1}{\sqrt{3}}\int_0^1 \frac{1}{(q-p x_2)^2} d{x_2} \int_0^{1-x_2}d{x_1}\big[\big(A_1-V_1\big) (x_1,x_2,1-x_1-x_2)\big]\nonumber\\
 &
 +\frac{ 1}{\sqrt{3}}\int_0^1 \frac{1}{(q-p x_3)^2} d{x_3} \int_0^{1-x_3}d{x_1}\big[\big(A_1-V_1-T_1\big)(x_1,1-x_1-x_3,x_3)\big]\nonumber\\
 &
+ \frac{ m_N^2}{\sqrt{3}}\int_0^1 \frac{(1+x_2)}{(q-p x_2)^4} d{x_2} \int_0^{1-x_2}d{x_1}\big[\big(-A_1^M+2V_1^M-2T_1^M\big) (x_1,x_2,1-x_1-x_2)\big]\nonumber\\
 &
 +\frac{ m_N^2}{\sqrt{3}}\int_0^1 \frac{(1+x_3)}{(q-p x_3)^4} d{x_3} \int_0^{1-x_3}d{x_1}\big[\big(-A_1^M+V_1^M-2T_1^M\big)(x_1,1-x_1-x_3,x_3)\big]\nonumber\\
 &
 +\frac{ m_N^2}{2\sqrt{3}}\int_0^1 \frac{(1+2y+y^2)}{(q-p y)^4} dy \int_y^1 d{x_2} \int_0^{1-x_2}d{x_1}\big[\big(2A_1-2A_2+2A_3-S_1+S_2-2V_1+2V_2\nonumber\\
 &+V_3+V_4\big)(x_1,x_2,1-x_1-x_2)\big]\nonumber\\
 &
 +\frac{ m_N^2}{2\sqrt{3}}\int_0^1 \frac{(1+2y+y^2)}{(q-p y)^4} dy \int_y^1 d{x_3} \int_0^{1-x_3}d{x_1}\big[\big(A_3-A_4-4T_1+4T_3+4T_7-2V_1+2V_3\nonumber\\
 &+2V_5\big)(x_1,1-x_1-x_3,x_3)\big]\nonumber\\
 &
 +\frac{2m_N^2}{\sqrt{3}}\int_0^1 \frac{(1+v)}{(q-p v)^4} dv \int_0^v dy \int_y^1 d{x_2} \int_0^{1-x_2} d{x_1} \big[\big(T_2-T_3-T_4+T_5+T_6+T_7+T_8\big)\nonumber\\
  & \times (x_1,x_2,1-x_1-x_2)\big]
 \nonumber\\
 &-\frac{2m_N^2}{\sqrt{3}}\int_0^1 \frac{(1+v)}{(q-p v)^4} dv \int_0^v dy \int_y^1 d{x_3} \int_0^{1-x_3} d{x_1}\big[\big(T_2-T_3-T_4+T_5+T_6+T_7+T_8\big)\nonumber\\
  & \times(x_1,1-x_1-x_3,x_3)\big],
 \\
  \nonumber\\
 \rho_6^{QCD}(v,y,x_1,x_2,x_3)&=\frac{m_N^2}{4\sqrt{3}}\int_0^1 \frac{1}{(q-p x_2)^2} d{x_2} \int_0^{1-x_2}d{x_1}\big[\big(A_1^M+2T_1^M-2V_1^M\big) (x_1,x_2,1-x_1-x_2)\big]\nonumber\\
 &
 +\frac{m_N^2}{4\sqrt{3}}\int_0^1 \frac{1}{(q-p x_3)^2} d{x_3} \int_0^{1-x_3}d{x_1}\big[\big(A_1^M-V_1^M\big)(x_1,1-x_1-x_3,x_3)\big]\nonumber
\end{align}

\begin{align}
&
 +\frac{m_N^2}{4\sqrt{3}}\int_0^1 \frac{(1+y)}{(q-p y)^2} dy \int_y^1 d{x_2} \int_0^{1-x_2}d{x_1} \big[\big(A_3-A_4-2P_1+2P_2+2S_1-2S_2+2T_1-2T_2  \nonumber\\
 & 
 -4T_7+V_3+V_4  \big)(x_1,x_2,1-x_1-x_2)\big]
 \nonumber\\
&+\frac{m_N^2}{2\sqrt{3}}\int_0^1 \frac{(1+y)}{(q-p y)^2} dy \int_y^1 d{x_3} \int_0^{1-x_3}d{x_1} \big[\big(A_1-A_2+A_4-2T_2+2T_3-2T_7-V_1+V_2 \nonumber\\
 &+ V_4\big)(x_1,1-x_1-x_3,x_3)\big]\nonumber\\
 &
 +\frac{m_N^2}{2\sqrt{3}}\int_0^1 \frac{1}{(q-p v)^2} dv \int_0^v dy \int_y^1 d{x_2} \int_0^{1-x_2} d{x_1} \big[\big(T_2-T_3-T_4+T_5+T_6+T_7+T_8\big)\nonumber\\
  & \times (x_1,x_2,1-x_1-x_2)\big]
 \nonumber\\
 &-\frac{m_N^2}{2\sqrt{3}}\int_0^1 \frac{1}{(q-p v)^2} dv \int_0^v dy \int_y^1 d{x_3} \int_0^{1-x_3} d{x_1}\big[\big(T_2-T_3-T_4+T_5+T_6+T_7+T_8\big)\nonumber\\
  & \times(x_1,1-x_1-x_3,x_3)\big]\nonumber\\
  &
  +\frac{m_N^4}{2\sqrt{3}}\int_0^1 \frac{v(1+v)}{(q-p v)^4} dv \int_0^v dy \int_y^1 d{x_2} \int_0^{1-x_2} d{x_1} \big[\big(T_1-T_2-T_5+T_6-2T_7-2T_8\big)\nonumber\\
  & \times (x_1,x_2,1-x_1-x_2)\big]
 \nonumber\\
 &-\frac{m_N^4}{2\sqrt{3}}\int_0^1 \frac{v(1+v)}{(q-p v)^4} dv \int_0^v dy \int_y^1 d{x_3} \int_0^{1-x_3} d{x_1}\big[\big(A_1-A_2+A_3+A_4-A_5+A_6+T_1+T_2\nonumber\\
  &-2T_3-2T_4+T_5-V_1+V_2+V_3+V_4+V_5-V_6 \big)(x_1,1-x_1-x_3,x_3)\big],
  \\
 \nonumber\\
  \rho_7^{QCD}(v,y,x_1,x_2,x_3)&=\frac{1}{\sqrt{3}}\int_0^1 \frac{(1+2x_2+x_2^2)}{(q-p x_2)^2} d{x_2} \int_0^{1-x_2}d{x_1}\big[\big(A_1-V_1\big) (x_1,x_2,1-x_1-x_2)\big]\nonumber\\
 &
 +\frac{2}{\sqrt{3}}\int_0^1 \frac{(1+2x_3+x_3^2)}{(q-p x_3)^2} d{x_3} \int_0^{1-x_3}d{x_1}\big[\big(V_1-A_1-T_1\big)(x_1,1-x_1-x_3,x_3)\big]\nonumber\\
 &
 +\frac{2m_N^2}{\sqrt{3}}\int_0^1 \frac{(1+2x_2+x_2^2)}{(q-p x_2)^4} d{x_2} \int_0^{1-x_2}d{x_1}\big[\big(A_1^M+T_1^M-V_1^M\big) (x_1,x_2,1-x_1-x_2)\big]\nonumber\\
 &
 +\frac{m_N^2}{\sqrt{3}}\int_0^1 \frac{(1+2x_3+x_3^2)}{(q-p x_3)^4} d{x_3} \int_0^{1-x_3}d{x_1}\big[\big(-7A_1^M-8T_1^M+7V_1^M\big)(x_1,1-x_1-x_3,x_3)\big]\nonumber\\
 &
  +\frac{4m_N^2}{\sqrt{3}}\int_0^1 \frac{v(1+v)}{(q-p v)^4} dv \int_0^v dy \int_y^1 d{x_2} \int_0^{1-x_2} d{x_1} \big[\big(T_2-T_3-T_4+T_5+T_6+T_7+T_8\big)\nonumber\\
  & \times (x_1,x_2,1-x_1-x_2)\big]
 \nonumber\\
 &-\frac{4m_N^2}{\sqrt{3}}\int_0^1 \frac{v(1+v)}{(q-p v)^4} dv \int_0^v dy \int_y^1 d{x_3} \int_0^{1-x_3} d{x_1}\big[\big(T_2-T_3-T_4+T_5+T_6+T_7+T_8 \big)\nonumber\\
  & \times(x_1,1-x_1-x_3,x_3)\big]\nonumber\\
  &
  +\frac{ m_N^2}{\sqrt{3}}\int_0^1 \frac{y(y+1)}{(q-p y)^4} dy \int_y^1 d{x_2} \int_0^{1-x_2}d{x_1}\big[\big(A_1-A_2+A_3-V_1+V_2+V_3\big)\nonumber\\
 & \times (x_1,x_2,1-x_1-x_2)\big]\nonumber\\
 &
 +\frac{ m_N^2}{\sqrt{3}}\int_0^1 \frac{y(y+1)}{(q-p y)^4} dy \int_y^1 d{x_3} \int_0^{1-x_3}d{x_1}\big[\big(A_3-A_4-2T_1+4T_2-2T_3+4T_5+2T_7+8T_8\nonumber\\
 &+2V_1-2V_4+2V_5\big)(x_1,1-x_1-x_3,x_3)\big]\nonumber\\
 &
 -\frac{4m_N^4}{ \sqrt{3}}\int_0^1 \frac{v^2(1+v)}{(q-p v)^6} dv \int_0^v dy \int_y^1 d{x_3} \int_0^{1-x_3} d{x_1}\big[\big(A_1-A_2+A_3+A_4-A_5+A_6+2T_1 \nonumber\\
 &-2T_3-2T_4+2T_6-2T_7-2T_8-V_1+V_2+V_3+V_5-V_6\big)(x_1,1-x_1-x_3,x_3)\big],\\
 \nonumber
%
\end{align}

\begin{align}
  \rho_8^{QCD}(v,y,x_1,x_2,x_3)&=\frac{m_N}{\sqrt{3}}\int_0^1 \frac{(1+2x_2+x_2^2)}{(q-p x_2)^2} d{x_2} \int_0^{1-x_2}d{x_1}\big[\big(A_1+A_3-V_1+V_2\big) (x_1,x_2,1-x_1-x_2)\big]\nonumber\\
 &
 +\frac{m_N}{2\sqrt{3}}\int_0^1 \frac{(1+2x_3+x_3^2)}{(q-p x_3)^2} d{x_3} \int_0^{1-x_3}d{x_1}\big[\big(-A_1-A_3-2T_1+V_1-2V_3\big)(x_1,1-x_1-x_3,x_3)\big]\nonumber\\
 &
 +\frac{m_N^3}{\sqrt{3}}\int_0^1 \frac{(1+2x_2+x_2^2)}{(q-p x_2)^4} d{x_2} \int_0^{1-x_2}d{x_1}\big[\big(-T_1^M-2V_1^M-A_1^M\big) (x_1,x_2,1-x_1-x_2)\big]\nonumber\\
 &
 +\frac{m_N^3}{2\sqrt{3}}\int_0^1 \frac{(1+2x_3+x_3^2)}{(q-p x_3)^4} d{x_3} \int_0^{1-x_3}d{x_1}\big[\big(A_1^M-V_1^M+T_1^M\big)(x_1,1-x_1-x_3,x_3)\big]\nonumber\\
  &  +\frac{ m_N^3}{2\sqrt{3}}\int_0^1 \frac{y(1+y+y^2)}{(q-p y)^4} dy \int_y^1 d{x_2} \int_0^{1-x_2}d{x_1}\big[\big(4A_1+4A_2-A_3-A_4+2S_1-2S_2+4V_1\nonumber\\
 &-4V_2-2V_3-2V_4-2V_5 \big) (x_1,x_2,1-x_1-x_2)\big]\nonumber\\
 &
 +\frac{ m_N^3}{2\sqrt{3}}\int_0^1 \frac{y(1+y+y^2)}{(q-p y)^4} dy \int_y^1 d{x_3} \int_0^{1-x_3}d{x_1}\big[\big(-A_3-2A_4+4T_1-4T_3-4T_7+2V_1-2V_3\nonumber\\
 &-2V_5\big)(x_1,1-x_1-x_3,x_3)\big]\nonumber\\
 &
 +\frac{m_N^3}{ \sqrt{3}}\int_0^1 \frac{(1+v+v^2)}{(q-p v)^4} dv \int_0^v dy \int_y^1 d{x_2} \int_0^{1-x_2} d{x_1}\big[\big(A_1-A_2+A_3+A_4-A_5+A_6+2T_1 \nonumber\\
 &-2T_3-2T_4+2T_6-2T_7-2T_8-V_1+V_2+V_3+V_5-V_6\big)(x_1,x_2,1-x_1-x_2)\big]\nonumber\\
 &
 +\frac{2m_N^3}{ \sqrt{3}}\int_0^1 \frac{(1+v+v^2)}{(q-p v)^4} dv \int_0^v dy \int_y^1 d{x_3} \int_0^{1-x_3} d{x_1}\big[\big(T_1-T_3-T_5+T_6-2T_7-2T_8\big) \nonumber\\
 & \times(x_1,1-x_1-x_3,x_3)\big],
 \end{align}
 and, 
 \begin{align}
 \rho_9^{QCD}(v,y,x_1,x_2,x_3)&=\frac{1}{\sqrt{3}}\int_0^1 \frac{(1+2x_2+x_2^2)}{(q-p x_2)^2} d{x_2} \int_0^{1-x_2}d{x_1}\big[\big(A_1-V_1\big) (x_1,x_2,1-x_1-x_2)\big]\nonumber\\
 &
 +\frac{2}{\sqrt{3}}\int_0^1 \frac{(1+2x_3+x_3^2)}{(q-p x_3)^2} d{x_3} \int_0^{1-x_3}d{x_1}\big[T_1(x_1,1-x_1-x_3,x_3)\big]\nonumber\\
 &
 +\frac{m_N^2}{2\sqrt{3}}\int_0^1 \frac{(1+2x_2+x_2^2)}{(q-p x_2)^4} d{x_2} \int_0^{1-x_2}d{x_1}\big[\big(-A_1^M+V_1^M\big) (x_1,x_2,1-x_1-x_2)\big]\nonumber\\
 &
 +\frac{m_N^2}{2\sqrt{3}}\int_0^1 \frac{(1+2x_3+x_3^2)}{(q-p x_3)^4} d{x_3} \int_0^{1-x_3}d{x_1}\big[T_1^M(x_1,1-x_1-x_3,x_3)\big]\nonumber\\
&
 +\frac{2m_N^2}{ \sqrt{3}}\int_0^1 \frac{(1+v+v^2)}{(q-p v)^4} dv \int_0^v dy \int_y^1 d{x_2} \int_0^{1-x_2} d{x_1}\big[\big(T_2+T_3+T_4-T_5-2T_7-2T_8\big) \nonumber\\
 &\times(x_1,x_2,1-x_1-x_2)\big]\nonumber\\
 &
 -\frac{2m_N^2}{ \sqrt{3}}\int_0^1 \frac{(1+v+v^2)}{(q-p v)^4} dv \int_0^v dy \int_y^1 d{x_3} \int_0^{1-x_3} d{x_1}\big[\big(T_2+T_3+T_4-T_5-2T_7-2T_8\big) \nonumber\\
 & \times(x_1,1-x_1-x_3,x_3)\big].
\end{align}

The  Borel transformation as well as the continuum subtraction are performed  making use of  the following replacements  \cite{Braun:2006hz}:
\begin{align}
		\int dx \frac{\rho(x)}{(q-xp)^2}&\rightarrow -\int_{x_0}^1\frac{dx}{x}\rho(x) e^{-s(x)/M^2}, \nonumber		\\
		\int dx \frac{\rho(x)}{(q-xp)^4}&\rightarrow \frac{1}{M^2} \int_{x_0}^1\frac{dx}{x^2}\rho(x) e^{-s(x)/M^2}
		+\frac{\rho(x_0)}{Q^2+x_0^2 m_{N}^2} e^{-s_0/M^2},\nonumber
		\end{align}
		\begin{align}
	\int dx \frac{\rho(x)}{(q-xp)^6}&\rightarrow -\frac{1}{2M^4}\int_{x_0}^1\frac{dx}{x^3}\rho(x) e^{-s(x)/M^2}
		-\frac{1}{2M^2}\frac{\rho(x_0)}{x_0(Q^2+x_0^2m_{N}^2)}e^{-s_0/M^2}\nonumber\\
		&+\frac{1}{2}\frac{x_0^2}{Q^2+x_0^2m_{N}^2}\bigg[\frac{d}{dx_0}\frac{\rho(x_0)}{x_0(Q^2+x_0^2m_{N}^2)}\bigg]e^{-s_0/M^2},
	\label{subtract3}
\end{align}
where
\begin{eqnarray}
s(x)=(1-x)m_{N}^2+\frac{1-x}{x}Q^2,
\end{eqnarray}
and $M^2$ is the Borel mass  parameter, which enters in the calculations after applying  the  Borel transformation  in terms of  $p^{\prime 2}$. Here,  $x_0$ is  found by solving the equation,  $s(x)=s_0$, i.e.,
\begin{eqnarray}
x_0&=&\Big[\sqrt{(Q^2+s_0-m_{N}^2)^2+4m_{N}^2 Q^2}-(Q^2+s_0-m_{N}^2)\Big]/2m_{N}^2,
\end{eqnarray}
with $s_0$ being the continuum threshold parameter.

\bibliography{NDeltaGFFs}
\end{document}